%% file: main_IUI2026.tex
\documentclass[sigconf]{acmart}
\AtBeginDocument{%
  }

\copyrightyear{2027}
\acmYear{2027}
\setcopyright{cc}
\setcctype{by-nc-nd}
\acmConference[IUI '27]{32nd International Conference on Intelligent User Interfaces}{February 08--11, 2027}{Helsinki, Finland}
\acmBooktitle{32nd International Conference on Intelligent User Interfaces (IUI '27), February 08--11, 2027, Helsinki, Finland}
\acmDOI{10.1145/3817122.3872804}
\acmISBN{979-8-4007-2765-8/2027/02}

\acmISBN{978-1-4503-XXXX-X/2018/06}

\usepackage{subcaption}

\usepackage{array}
\usepackage{ragged2e}
\usepackage{booktabs}
\usepackage{colortbl}

\newcolumntype{C}[1]{>{\centering\arraybackslash}p{#1}}

\author{Wazeer Zulfikar}
\authornote{Contributed equally to this research.}
\email{wazeer@media.mit.edu}
\orcid{0009-0008-7753-8817}
\affiliation{%
  \institution{MIT Media Lab}
  \city{Cambridge}
  \country{USA}
}

\author{Sally Ahmed}
\authornotemark[1]
\email{sallyelf@media.mit.edu}
\affiliation{%
  \institution{MIT Media Lab}
  \city{Cambridge}
  \country{USA}
}

\author{Manasi Vaidya}
\email{manasiv@media.mit.edu}
\affiliation{%
  \institution{MIT Media Lab}
  \city{Cambridge}
  \country{USA}
}
\additionalaffiliation{%
  \institution{MIT AgeLab}
  \city{Cambridge}
  \country{USA}
}

\author{Samarth KJ}
\email{ce22b099@alumni.iitm.ac.in}
\affiliation{%
  \institution{Indian Institute of Technology}
  \city{Madras}
  \country{India}
}

\author{Pattie Maes}
\email{pattie@media.mit.edu}
\orcid{0000-0002-7722-6038}
\affiliation{%
  \institution{MIT Media Lab}
  \city{Cambridge}
  \country{USA}
}

\newenvironment{heroquote}
  {\begin{quote}\itshape}
  {\end{quote}}
  
\begin{document}


\title{Always-On, But Worth It? How Older Adults Use a Wearable AI Memory Agent in Daily Life}



\renewcommand{\shortauthors}{Zulfikar \& Ahmed et al.}


\begin{abstract}

Always-on wearable AI can support older adults' memory, but its continuous presence raises the question of when such support is worth its social and behavioral costs. We deployed \textsc{Memoro 2}, an always-on, audio-based AI memory agent, as a technology probe with 10 older adults aged 60 to 86 for 10 days in their own contexts. \textsc{Memoro 2} supports memory through on-demand recall of past conversations, proactive reminders extracted from speech, and end-of-day summaries. Across 144 hours of transcribed conversation, 464 memory interactions, and 60 interviews, the probe surfaced three findings. (i) Adoption was a context-dependent tradeoff that followed daily rhythms: reminders cued morning routines and recalls served as end-of-day review. In the situations participants chose to use it, 96\% of recalls were rated helpful, while in the situations where support would have mattered most, participants chose not to use it at all. Skips concentrated in high-stakes settings and sensitive conversations where recording carried legal and social costs. (ii) Although designed as a personal aid, \textsc{Memoro 2} functioned as a shared social tool. Reviewing summaries and recalls prompted reflection on communication styles, and the record drifted into a joint utility for settling disagreements and maintaining relationships. (iii) Participants wanted always-on capture they could regulate in the moment: adjusting how visibly the system gives feedback to fit the setting, treating bystander consent as ongoing, and curating what the record retains so capturing everything stays manageable. These findings suggest that the value of always-on memory support extends beyond the individual user and depends on older adults' ability to control it in the moment.

\end{abstract}


\begin{teaserfigure}
  \includegraphics[width=\textwidth]{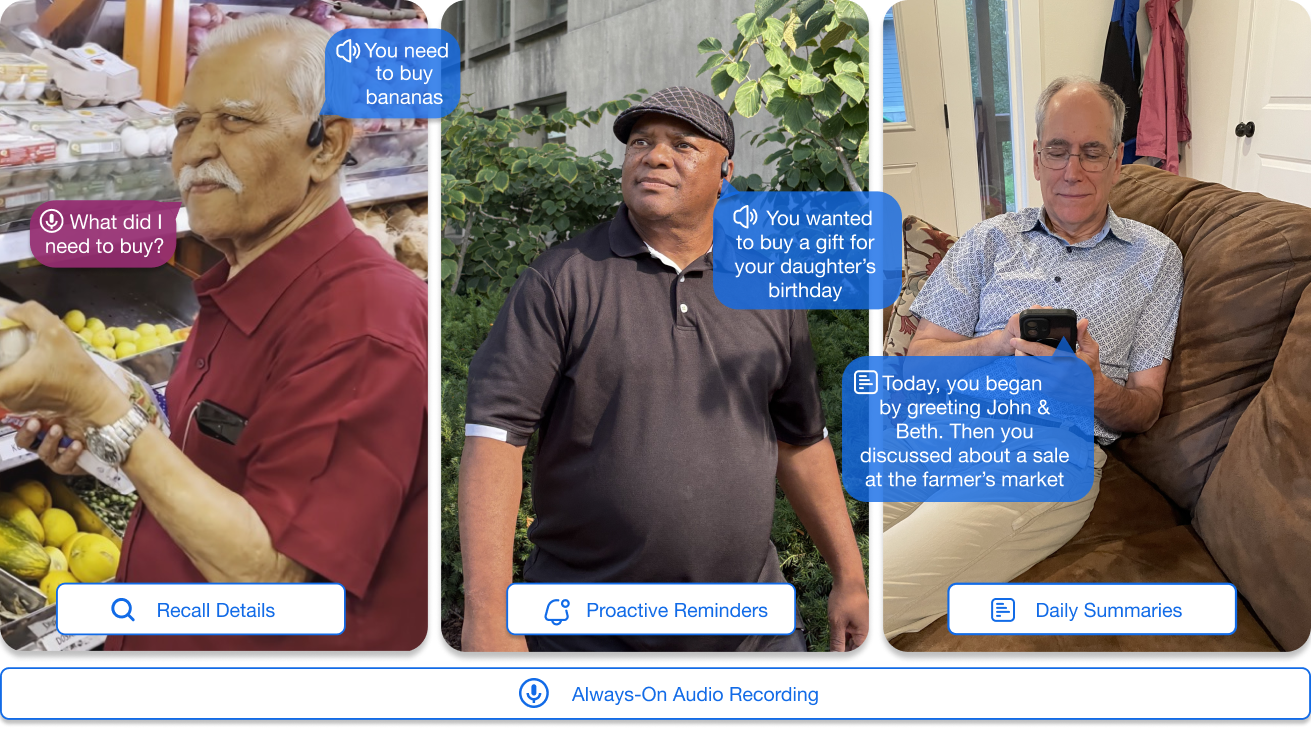}
  \caption{\textsc{Memoro 2}, an audio-based wearable AI memory agent. The system continuously records audio and stores conversations as memories. It supports human memory in three ways (left to right): (i) voice-based recall of past conversation details, answered as audio through the headset; (ii) proactive reminders of commitments spoken in conversation, delivered as audio and phone notifications; and (iii) end-of-day summaries of the day's conversations, shown on the phone.}
  \Description{Example scenarios depict recalls, proactive reminders, and daily summaries on the phone UI, conveying the envisioned user value of the always-on memory assistant with three older adults.}
  \label{fig:teaser}
\end{teaserfigure}

\received{20 February 2007}
\received[revised]{12 March 2009}
\received[accepted]{5 June 2009}

\maketitle

\section{Introduction}

Older adults manage appointments, household tasks, relationships, and community roles whose upkeep supports functional independence and wellbeing \cite{godin_impact_2019, stites_relationships_2018, kliegel2016prospective, shaikh_impact_2021}. However, as everyday demands accumulate, growing lapses in memory strain this upkeep \cite{ossher_everyday_2013, stites_relationships_2018}. Such lapses are unpredictable, so support cannot wait for users to anticipate what they will forget. An always-on wearable system counters this by capturing first-person experiences as they happen and resurfacing them when memory falters through just-in-time support \cite{franklin2021designing, rhodes_just--time_2000, vemuri_audio-based_2004, shah_lifelogging_2012, rhodes_remembrance_1996}.

Advances in speech and language models, combined with low-power, lightweight wearables, make the long-standing promise of always-on memory support feasible in daily life \cite{bush1945we, chen_llamapie_2025, zulfikar_memoro_2024, fang_mirai_2025}. Specifically for older adults, prior HCI work on memory augmentation has explored memory support in several ways such as prospective memory training \cite{chan_prospecfit_2019}, object tracking \cite{li_fmt_2019}, and safety reminders \cite{maniar_mempal_2025}. Each targets a single memory function and showcases promising capabilities. Such systems are always-on in two senses that distinguish them from wake-word voice assistants: they continuously capture and retain speech, and they are continuously ready to deliver reminders, summaries, and on-demand recall. However, successful technology adoption further depends on how it integrates into daily contexts. The continuous presence of always-on systems introduces social and ethical complexities beyond the system's technical challenges, which can reshape individual behavior and interpersonal relations \cite{denning2014bystanders, hoyle_privacy_2014, wang2026mindgap, ohagan2023pet}. Further, technology adoption among older adults hinges on whether a tool provides recurring value, fits established routines, and avoids marking its user as declining \cite{Czaja2006Factors, lee2015olderAdultsAdoption, caldeira2022stigma, knowles2021harm, chan_co_designing_2024}. For older adults, then, always-on support is not a one-time adoption decision but a continuing tradeoff between recurring memory value and the social and behavioral costs of being recorded. How always-on memory support is taken up within the social ecology of older adults' daily life remains unexplored. Therefore, we investigate the following research questions:


\begin{itemize}
    \item \textbf{RQ1.} How do older adults take up and integrate an always-on, audio-based AI memory agent into their daily routines?
    \item \textbf{RQ2.}  How does the continuous presence of the system shape the social behavior of users and bystanders during interpersonal interactions?
    \item \textbf{RQ3.} What design considerations for future always-on memory support emerge from older adults' lived experience of using such a system?
\end{itemize}


To investigate these questions, we built \textsc{Memoro 2} \footnote{The \textsc{Memoro 2} mobile application code will be released as open source upon publication; the repository link is withheld for review anonymity.} (See Fig. \ref{fig:teaser}), an always-on, audio-based AI memory agent used as a technology probe \cite{hutchinson_technology_2003} to surface needs that emerge through use. \textsc{Memoro 2} extends Memoro \cite{zulfikar_memoro_2024}, a wearable system that answers voice queries about past conversations, with two proactive capabilities: automated commitment extraction and daily summaries, enabling prospective memory support. These additions were informed by a needs-finding survey of N=100 older adults. To support continuous capture in naturalistic environments and address privacy concerns among older adults \cite{yusif2016older, bonilla2020older}, we implemented a Privacy-by-Design architecture \cite{cavoukian2009privacy} with local encryption, automated audio deletion post-transcription, a dashboard for quick deletion, and zero-data-retention processing. \textsc{Memoro 2} uses a bone-conduction headset that leaves hearing unoccluded for natural conversation and, importantly, is a generic commercial form factor rather than a marked assistive device, addressing the comfort and assistive-stigma factors in technology adoption among older adults. 

We deployed \textsc{Memoro 2} in a 10-day free-living field study with 10 participants aged 60 to 86, capturing 144 hours of transcribed conversation \footnote{Captured hours measure usable, transcribed conversational speech retained after Voice Activity Detection (VAD) filtering; they exclude silence, ambient sound, and time spent not in conversation.}, 464 memory interactions, and 60 interviews, and evaluated its outputs through privacy-preserving in-situ ratings paired with offline validation. We make three contributions:

\begin{enumerate}
    \item The first multi-day, in-the-wild deployment of an always-on wearable AI memory agent with older adults, evaluated through privacy-preserving in-situ ratings and offline validation, showing adoption as a context-dependent tradeoff between memory value and the social costs of recording.
    \item Evidence of always-on memory support functioning as a \textit{shared social tool}, prompting reflection on one's communication styles and drifting into a joint utility for settling disagreements and maintaining relationships.
    \item Design directions for \textit{adaptively-on use} through always-on capture, drawn from older adults' suggestions such as adjustable visibility, post-capture controls, and ongoing consent management features.
\end{enumerate}
    
\section{Related Work}
Our work is related to and inspired by past work on digital memory augmentation systems, voice assistant studies among older adults, and the effects of always-on AI systems on users.

\subsection{Memory Augmentation Systems}

Since Vannevar Bush’s 1945 vision of the Memex, a conceptual digital memory assistant, successive waves of technologies have driven tool development for augmenting human memory \cite{bush1945we}. Digital archives such as MyLifeBits enabled searching through logs of past information \cite{karger_haystack_2004, gemmell_mylifebits_2002}. Later, wearable systems such as SenseCam and the Personal Audio Loop (PAL) allowed people to continuously capture daily experiences in the form of lifelogs \cite{hutchison_sensecam_2006, vemuri_audio-based_2004, lamming_design_1994, devaul_memory_2004,hutchison_personal_2004, shah_lifelogging_2012, harvey_remembering_2016}. Projects such as VIMES and DeepDiary enabled semantic searching through these lifelogs using high-level semantic concepts \cite{bermejo_vimes_2020, fan_deepdiary_2018}. Conversational AI then gave rise to systems such as Memoro and Llamapie, which treat memory assistance as a dialogue allowing natural language interactions \cite{zulfikar_memoro_2024, maniar_mempal_2025, chen_llamapie_2025, cai_ubiquitous_2025, li2025omniquery}. Together, they illustrate a progression from passive data logging to interactive memory partners. Functionally, these systems fall into two categories (i) retrospective, recall of past events and information aiding in reflecting, revisiting, and reminiscing past experiences \cite{hutchison_sensecam_2006, gupta_total_2022, cai_ubiquitous_2025, zulfikar2025resonance, zulfikar2026mnemonic}, and (ii) prospective, remembering to act on future intentions, assisting in future tasks, intentions, and promises \cite{chan_prospero_2019, chan_prospecfit_2019}. Both types of assistance can support a diverse range of users, while older adults particularly benefiting due to age-related cognitive decline.

Although prior studies have explored memory augmentation with older adults, the in-the-wild exploration of adoption remains underexplored. ProspecFit digitized a lab-based prospective-memory intervention, showing 12-day improvements with 10 older adults \cite{chan_prospecfit_2019}. FMT used fiducial markers and wearable cameras to track household objects, demonstrating feasibility but limited daily-life coverage \cite{li_fmt_2019}. MemPal applied multimodal models for object finding and safety reminders in short trials \cite{maniar_mempal_2025}. PENCODER targeted prospective-memory encoding for older adults with mild cognitive impairment constrained in a design-thinking lab phase \cite{shin2021pencoder}. These systems solely address either retrospective or prospective memory support and rely on short, controlled, or narrowly scoped deployments. To our knowledge, this is the first multi-day in-the-wild deployment of an always-on, wearable AI memory agent for older adults that integrates both retrospective recall and prospective support.

\subsection{Voice Assistants and Older Adults}

Voice-based assistants are increasingly studied in the context of aging, probing how they are taken up and fit into daily routines \cite{du2024enhancing, kim2021chb, mathur2025feels, pradhan2020iva, trajkova2020alexa, soubutts2022amazon}.
Early exposure studies emphasize that first impressions shape acceptability through simplicity and low-effort access \cite{kim2021firstinteractions}. Longitudinal deployments report selective, task-focused engagement that evolves with use \cite{chang2024longterm, kim2021chb};
Bentley et al.'s 88-home, 110-day deployment shows usage stabilizing around a small set of routines rather than expanding into the long tail of capabilities \cite{bentley_understanding_2018}, and Upadhyay et al.\ find that older adults recalibrate use over time as functional value becomes visible \cite{upadhyay2023studying}. Smart-speaker studies have also examined effects on loneliness \cite{jones2021lonelinessPVA}.
Privacy concerns among older adults center on ``listening,'' data sharing, and control, even as many accept convenience despite risk \cite{bonilla2020older, lau_alexa_2018, spangler2021privacy, yusif2016older, hoyle_privacy_2014, dunbar2021listening}; older adults often report lower day-to-day concern than younger adults yet still call for stronger protections and clearer controls over what is captured and retained \cite{spangler2021privacy}. Similarly, technology adoption among older adults is entangled with autonomy: devices that signal decline or invite external judgment are resisted \cite{caldeira2022stigma, kottl2021but}, and HCI work on aging that frames the user through a deficit lens has been critiqued for obscuring what older adults value: continued capability, discretion, and reciprocity in relationships \cite{knowles2021harm, vines2015ageold}. For wearable or memory-focused assistants, the same concerns extend to shared spaces and bystanders \cite{denning2014bystanders, kandappu2021privacyprimer, li2023privacy}.

Research on voice assistants for older adults has concentrated on domestic deployments, leaving mobility and out-of-home social contexts—where memory demands often surface—largely unstudied. Recent LLM-based agents tailored for older adults demonstrate richer interaction through supporting interruptions and backchannels \cite{liu2025natural, yang2024talk2care}, and a systematic review of conversational AI for aging reports that older adults prioritize naturalness, patience, and personalization over feature breadth \cite{huang2025designing}. We extend this line by deploying a wearable conversational system into the daily lives of older adults, examining how it operates across different contexts.

\subsection{Effects of Always-On AI Systems on Users}

Always-on AI systems are continuously present, available, and potentially active without explicit user input, creating new opportunities for support and new challenges for everyday life. Constant availability can reduce interaction friction and enable just-in-time, all-day assistance, supporting moments of need that are difficult to anticipate \cite{rhodes_just--time_2000, horvitz_principles_1999, chen_llamapie_2025}. For older adults, this means memory support can be woven seamlessly into daily routines, from recalling conversational details to setting reminders in situ. At the same time, continuous access also raises the risk of over-reliance, where users defer to the system in contexts where remembering or problem-solving might otherwise occur independently \cite{zhai_effects_2024}. Because always-on systems are also always recording, they create an ambient layer of data capture that can enable powerful retrospective and prospective assistance \cite{vemuri_audio-based_2004, hutchison_sensecam_2006, chan_prospero_2019}, but simultaneously highlight privacy and ethical concerns. Persistent recording introduces negotiations not only for the primary user but also for bystanders \cite{sehrt_social_2022, zulfikar_memic_2024}. These negotiations can affect adoption thresholds and social acceptability, particularly where recording is perceived as intrusive or stigmatizing, such as hospitals.

Always-on AI systems can also be proactive, surfacing reminders, suggestions, or summaries without being asked. Although proactivity may improve timeliness and reduce user burden \cite{hyland_what_2018, chen_llamapie_2025, fang_mirai_2025}, it also risks disrupting activities and conversations, misjudging relevance, or undermining the user's sense of autonomy \cite{fang_goldilocks_2025}. Recent proactive assistants on smart glasses act on their own initiative in short, session-based studies \cite{li2025satori, yang2025socialmind}, and studies of proactive dialogue tie how much initiative a system takes to how much users trust it \cite{kraus2021trust}. For older adults in particular, such trade-offs are entangled with questions of independence, trust, and social identity. In an in-the-wild study, \citet{meurisch_exploring_2020} found that although participants expected proactive support in areas such as social scheduling, they often rejected it in sensitive domains like mental health, citing intrusiveness and privacy concerns. Finally, there are the psychological and behavioral effects of an ever-present system. \citet{hoyle_privacy_2014} showed that the awareness of being observed or recorded can sometimes alter behavior. At the same time, these effects can be constructive: always-on systems may encourage accountability and goal-setting \cite{li_understanding_2011}. This dual role, support tool and social actor, motivates our deployment of an always-on wearable memory agent with older adults, examining both the perceived benefits of constant availability and the lived consequences of continuous presence.

\section{System Design and Development}

\textsc{Memoro 2} extends a pure conversational recall tool (Memoro) \cite{zulfikar_memoro_2024} into a multi-faceted memory agent, grounded in a preliminary need-finding phase to align its capabilities with the daily needs of older adults.

\subsection{Design Requirements}

To ensure the system addressed the most pressing memory challenges of aging we conducted a preliminary online survey to elicit functional requirements. The survey was exempt by the university ethics review board.

\subsubsection{Need-Finding Methodology and Participants}

The survey was hosted on CloudResearch and targeted 100 respondents over the age of 60 ($\mu_{age}=68.8, \sigma_{age}=5.8$) based in the United States of America, 85.7\% of whom were familiar with voice or AI assistants.

The survey followed a four-part protocol:

\begin{enumerate}
    \item System Orientation: Participants viewed a video demonstration of the original \textsc{Memoro} prototype (video obtained from the authors) to understand its baseline conversational capture capabilities \cite{zulfikar_memoro_2024}.
    \item Quantitative Ranking: Respondents ranked five candidate purposes for a memory aid (1 = most likely to use, 5 = least likely)
    \item Qualitative Scenarios: Participants were presented with different scenarios (How would you ask the tool to tell you about what you did yesterday?). 
    \item  Open-Ended Feedback: Participants articulated anticipated use cases for the tool.
    \end{enumerate}
The survey questions are in Appendix~\ref{appendix:meta:surveys}.

\subsubsection{From Preferences to Requirements}

\begin{figure*}[htbp]
  \centering
  \begin{subfigure}[b]{0.49\linewidth}
    \centering
    \vspace{20pt} 
    \includegraphics[width=\linewidth]{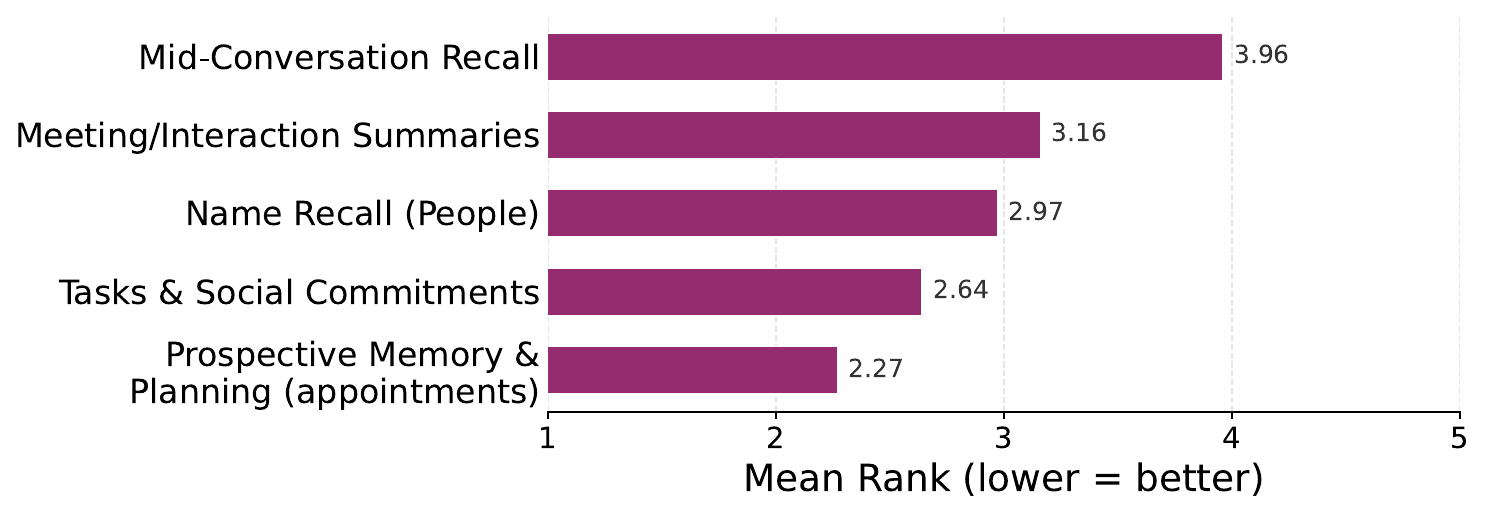}
    \vspace{20pt} 
    \caption{Most preferred purposes of using the tool (lower is better)}
    \label{fig:survey_purposes}
  \end{subfigure}
  \hfill
  \begin{subfigure}[b]{0.49\linewidth}
    \centering
    \includegraphics[width=\linewidth]{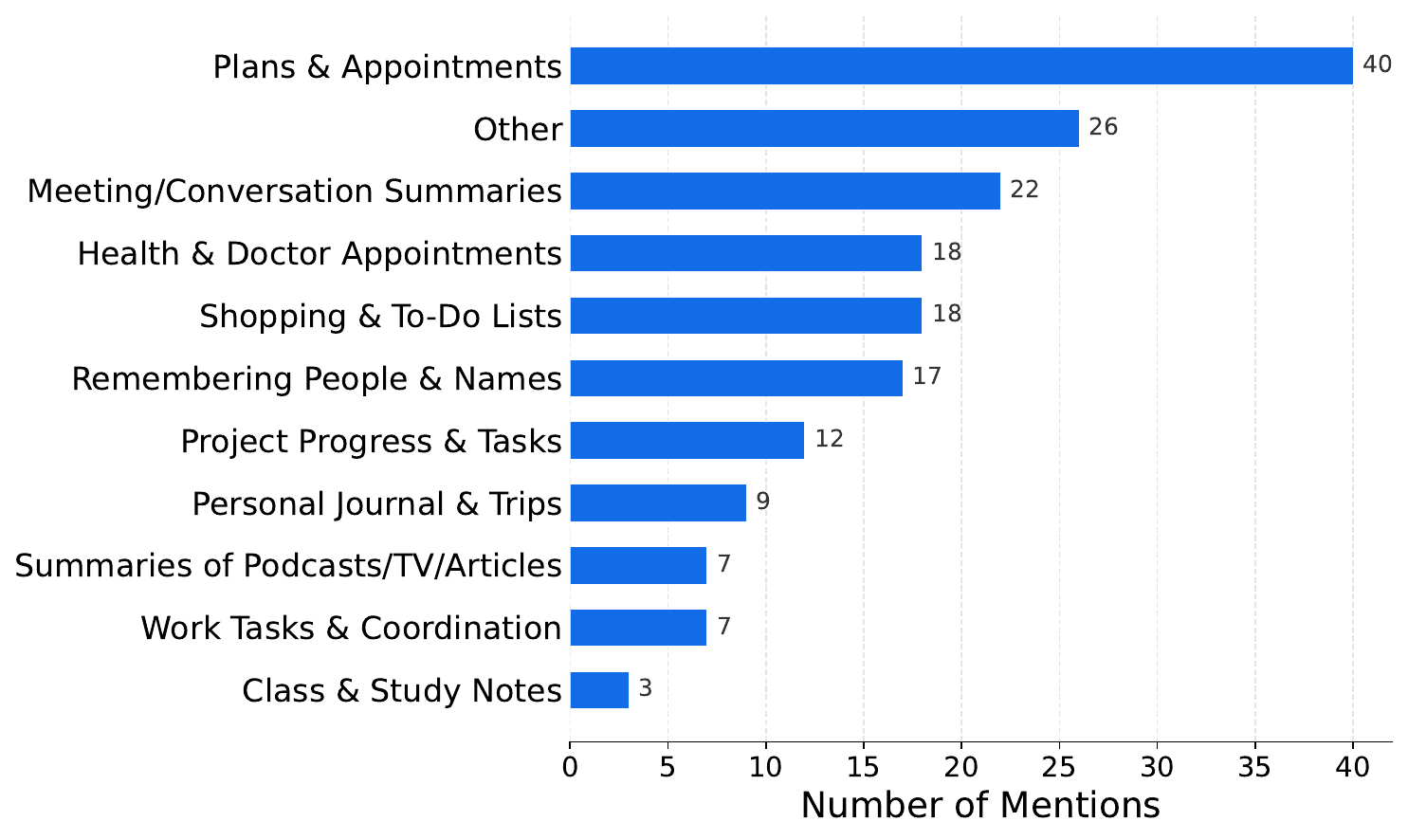}
    \caption{Survey recall themes}
    \label{fig:survey_themes}
  \end{subfigure}

  \caption{Preliminary survey results: (a) participants’ most preferred purposes of using the tool and (b) key themes in recall.}
  \Description{Preliminary survey results. Charts indicate participants most preferred “Prospective Memory \& Planning (appointments),” with tasks/commitments next; free-response themes similarly emphasize plans and appointments.}
  \label{fig:survey_results}
\end{figure*}

The results of the ranking task and free-response analysis (see Fig. \ref{fig:survey_results}) directly informed the gap in desired functionality through such a system:

\begin{itemize}
\item \textbf{Prioritization of Prospective Memory}: "Prospective Memory and Planning" (e.g., appointments) emerged as the highest priority, with a mean rank of $2.27$ and $30.8\%$ of participants placing it first.
\item \textbf{Demand for Task and Social Management}: The second-most preferred use case was "Tasks and Social Commitments" (doctor’s appointments, upcoming social events, birthdays, etc.), followed by "Meeting and Interaction Summary" (work meetings, conversations, and similar interactions).
\item \textbf{De-prioritization of Real-Time Recall}: "Mid-Conversation Recall" which was one of the focuses of the original system ranked lowest among the five categories.
\item \textbf{Specific Use Case Mentions}: Free-response analysis yielded $n=40$ mentions for "Plans and Appointments," $n=22$ for "Meeting/Conversation Summaries," and $n=18$ for "Health and Doctor Appointments".
\end{itemize}

These findings led to the formulation of two additional design requirements for \textsc{Memoro 2} in addition to the recall functionality offered by \textsc{Memoro} \cite{zulfikar_memoro_2024}:

\begin{enumerate}
    \item \textbf{Proactive Task Extraction}: The system must automatically detect actionable commitments from conversational audio without explicit user prompting.
    \item \textbf{Multi-Conversation Summarization}: The system must help review past conversations of the day through glanceable summaries.
\end{enumerate}

\subsection{System Architecture}

\begin{figure*}[h]
  \centering
    \includegraphics[width=\linewidth]{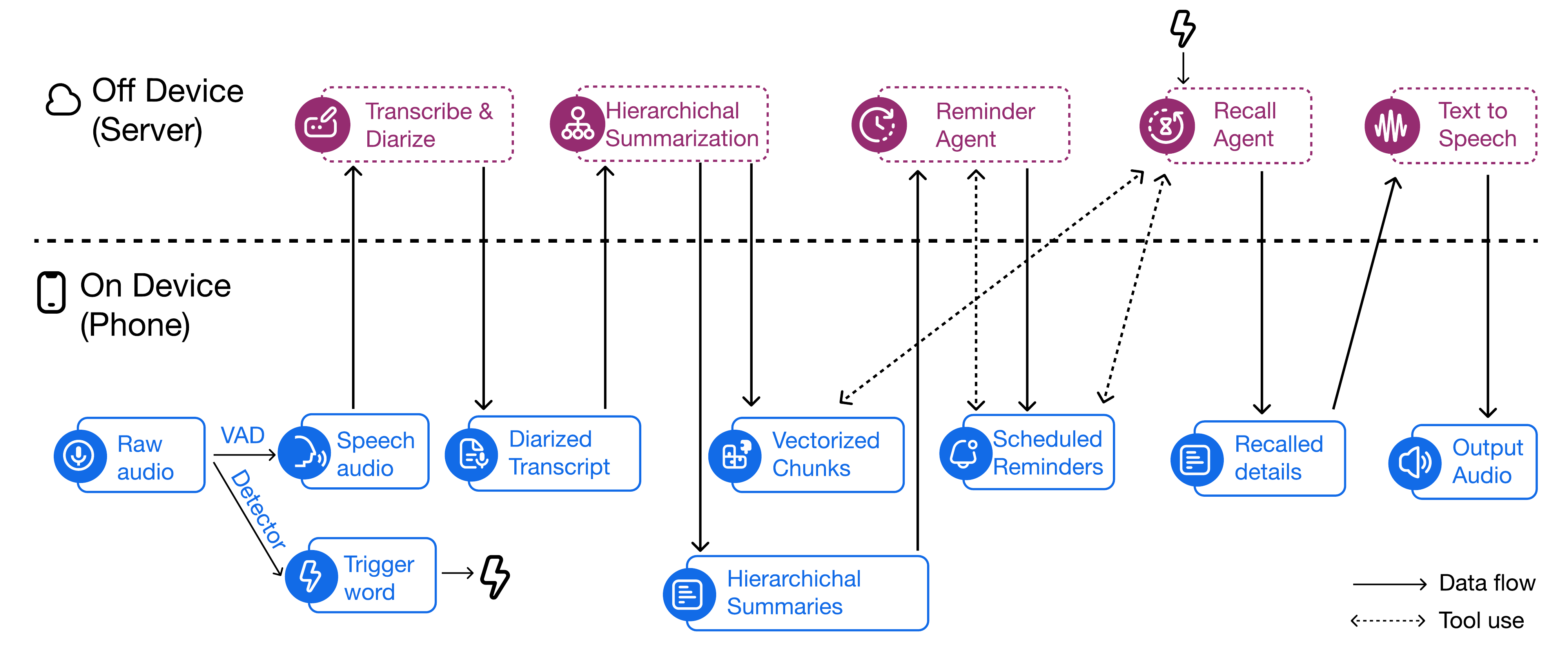}
    \caption{Overall system overview, showing the data flow and services used. All user data is stored locally and encrypted, while the services run either on or off device to maintain privacy and conserve bandwidth.}
  
  \Description{System architecture. End-to-end dataflow separates on-device components (VAD, trigger, local summaries) from off-device services (transcription/diarization, summarization, reminder and recall agents, text-to-speech), highlighting privacy-aware processing.}
  \label{fig:system_overview}
\end{figure*}

\textsc{Memoro 2} is an always-on, audio-based wearable AI agent comprising a bone conduction headset and a smartphone processing pipeline. Once activated, the system runs continuously in the background to reduce the cognitive burden of manual recording, automatically transforming detected speech into metadata-enriched digital memories stored locally for privacy. This continuous stream captures two primary verbal contexts: (i) \textbf{interactions} where users could be speaking aloud by themselves (self-talk) for \textsc{Memoro 2} to take note of, such as ``I should water the plants later'' and recall where users use the trigger word ``Hey Navi'' to ask questions to the system, such as "Hey Navi, when did Bob say we should meet?"; and (ii) \textbf{conversations} with other people while wearing \textsc{Memoro 2}, such as ``Hey John, let's meet on Friday for golfing''. The full system overview is illustrated in Fig. \ref{fig:system_overview}.

\subsubsection{Audio Capture}

\textsc{Memoro 2} continuously streams audio from the bone conduction headset’s microphone to the smartphone. To conserve bandwidth and server compute, an on-device Voice Activity Detection (VAD) monitors the stream and periodically sends 2–3 minute clips to the server only when speech has been detected. The pipeline operates through four states: (i) idle: a rolling 5-second buffer preserves context before an utterance; (ii) initiation: when speech is detected, an interaction/conversation starts and audio is prepended to a rolling buffer, marking onset; (iii) segmentation: recording continues up to 2 minutes, after which that clip is sent for transcription while a new clip immediately starts recording. The audio is segmented into clips in order to avoid heavy processing of single clips; and (iv) timeout: if silence persists for 30 seconds after 2 minutes, the interaction/conversation is closed, uploaded, and the system returns to idle with the buffer. These parameters are configurable defaults, tuned in pilot iteration to balance latency, bandwidth, and utterance coverage.

\subsubsection{Privacy and Data Security}\label{subsubsec:privacy} Given the sensitivity of continuous audio capture and known privacy concerns \cite{li2023privacy}, the architecture instantiates Cavoukian's foundational Privacy-By-Design principles \cite{cavoukian2009privacy} of proactive default protections, end-to-end security, full-lifecycle data minimization, and visible user control. The four mechanisms below realize these principles in deployment: \begin{itemize} 
\item \textbf{Encryption}: All user data is encrypted at rest (AES-256-GCM, device-keystore keys)
\item \textbf{Data Minimization}: Raw audio is processed on a secure server solely for transcription and is deleted immediately afterward from all systems. 
\item \textbf{Processing Protocols}: Transcription, diarization, and language-model tasks (OpenAI) operated under contractually enforced zero-data-retention agreements active for the duration of the study. No audio or transcript content was retained, logged for abuse monitoring, or used for model training by the provider. Audio uploaded for transcription was deleted on receipt of the transcript; raw audio was never persisted server-side.
\item \textbf{User Agency for Deletion}: The system provides an intuitive interface allowing users to easily and permanently delete any stored data from the device at any time, including erasing all data with a single button.
\end{itemize}

These mechanisms minimize retention and researcher access; they do not eliminate off-device processing, since raw audio transits commercial APIs under zero-retention contracts before deletion.


\subsection{Memory Encoding and Processing}

The raw audio captured by the hardware must be transformed into a structured, searchable knowledge base; this process begins with separating speakers, identifying the user and enriching the transcripts with temporal and social metadata.

\subsubsection{Diarization and Metadata Enrichment}

When an audio clip reaches the server, it is transcribed, diarized, and enriched with metadata to support the recall and reminder agents. During transcription, speaker labels are assigned through a dynamic reference set of voice signatures. The set is initialized with the primary user’s voice and expands as new speakers are detected within a conversation, i.e. there is no match with a speaker in the reference set, and is reset for every new conversation to prioritize privacy. When a new speaker is detected, 5–10 seconds of their speech is used to create a signature for that speaker within that conversation. While the system maintains diarization continuity within a single conversation, it does not perform cross-conversation speaker identification, since the accuracy was unreliable across environmental conditions. Cross-conversation queries about specific non-primary speakers therefore rely on names mentioned within transcripts (e.g., "thanks, Sarah") rather than voice signatures. Transcription and diarization were done using ElevenLabs API \cite{ElevenLabs}. The memory is then stored with metadata, including Conversation IDs, timestamps, and unique speaker counts.

\subsubsection{Hierarchical Summarization Pipeline} 
\begin{figure*}[h]
  \centering
    \includegraphics[width=0.7\linewidth]{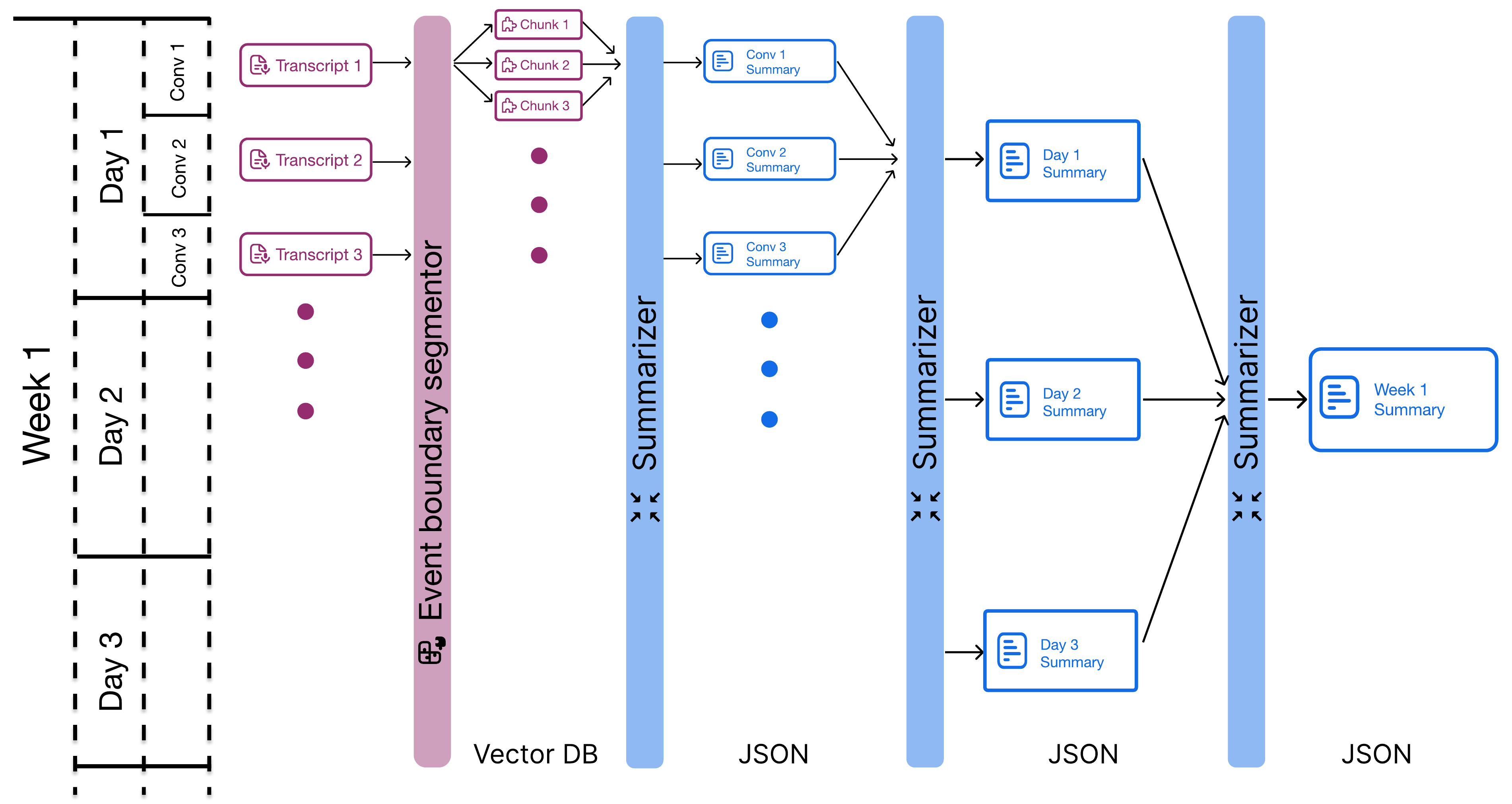}
    \caption{Stages of hierarchical summarization while processing each conversation transcript to extract semantic chunks, followed by summarizations at different temporal levels}
  
  \Description{Stages of hierarchical summary. Diagram shows transcripts segmented and summarized at multiple temporal levels (chunk → daily → weekly), stored with IDs and embeddings, enabling efficient retrieval and downstream agents.}
  \label{fig:hierarchical_memory}
\end{figure*}

To facilitate memory support across multiple days while bypassing LLM context window limits, the system employs a hierarchical summarization pipeline. As detailed in Fig. \ref{fig:hierarchical_memory}, transcripts undergo the following automated processing stages:

\paragraph{Chunk-Level Encoding} Conversation transcripts are segmented using GPT-4o into 3--8 coherent chunks per hour, separated by topic changes. For each segment, the system generates a 150--200 word summary that resolves pronouns and preserves specific details such as names, dates, and locations. These summaries are then vectorized using \textit{text-embedding-3-small} to enable semantic similarity search and assigned a significance score (0--10) via prompting GPT-4o to prioritize memories with high estimated future value. The prompts used for chunking, summarizing, and determining the significance are provided in the supplementary materials.

\paragraph{Semantic Aggregation} Summaries are recursively consolidated to maintain chronology and surface forward-looking details like plans, goals, and commitments. The hierarchy includes: \begin{itemize} \item \textbf{Conversation}: A 100-word summary and identified topic. \item \textbf{Daily}: A 200-word review of the day's interactions. \item \textbf{Weekly}: A 300-word high-level summary. \end{itemize}

\paragraph{Storage and UI Integration} All logs and chunks are stored locally with hierarchical identifiers, including parent IDs, chunk indices, and timestamps, alongside embeddings and metadata. Word limits are configurable defaults, set to fit the dashboard at accessible font sizes for older adults.

\subsubsection{Dashboard for Memory Review and Deletion}

\begin{figure*}[htbp]
  \centering
  \begin{subfigure}{0.19\linewidth}
    \includegraphics[width=\linewidth]{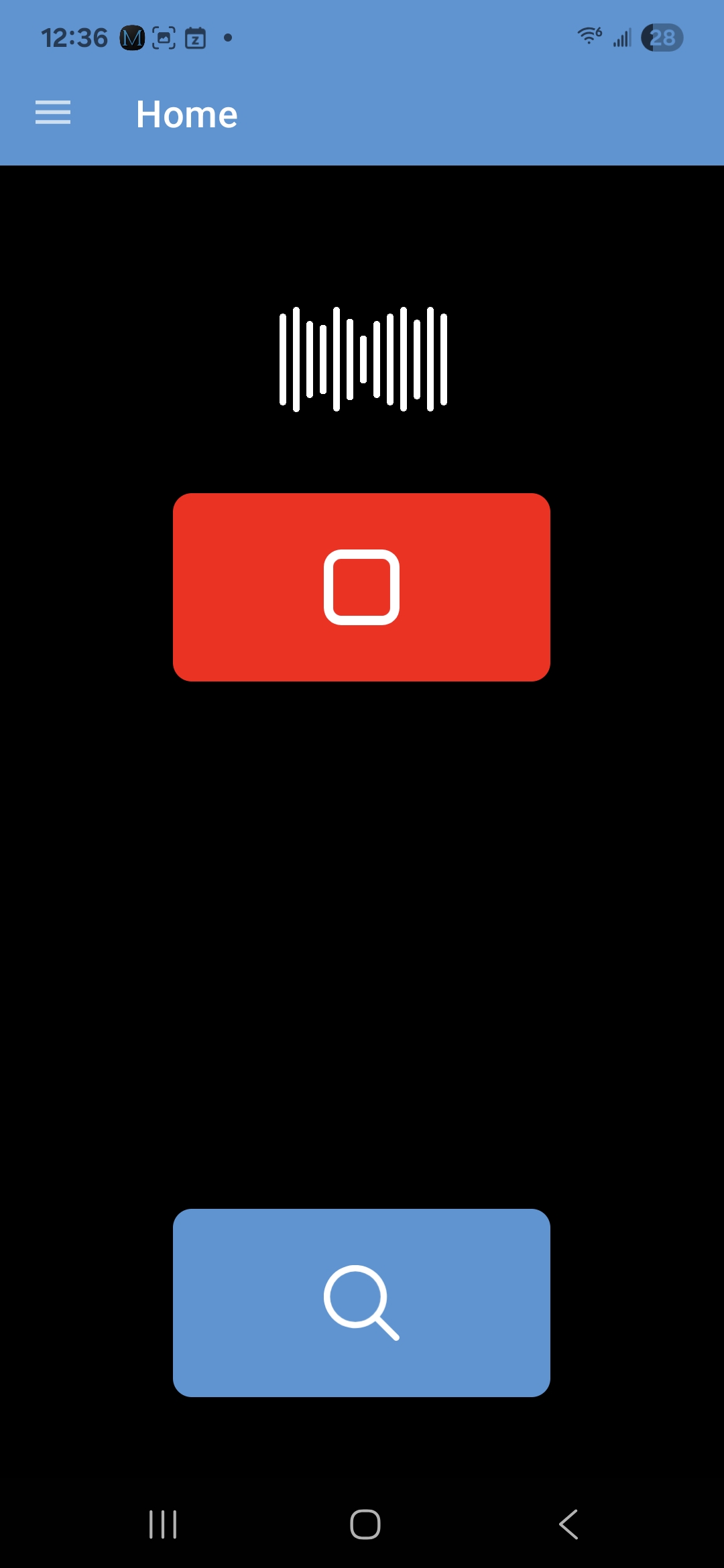}
    \caption{Recording Screen}
    \label{fig:screenshot_recording}
  \end{subfigure}
  \hfill
  \begin{subfigure}{0.19\linewidth}
    \includegraphics[width=\linewidth]{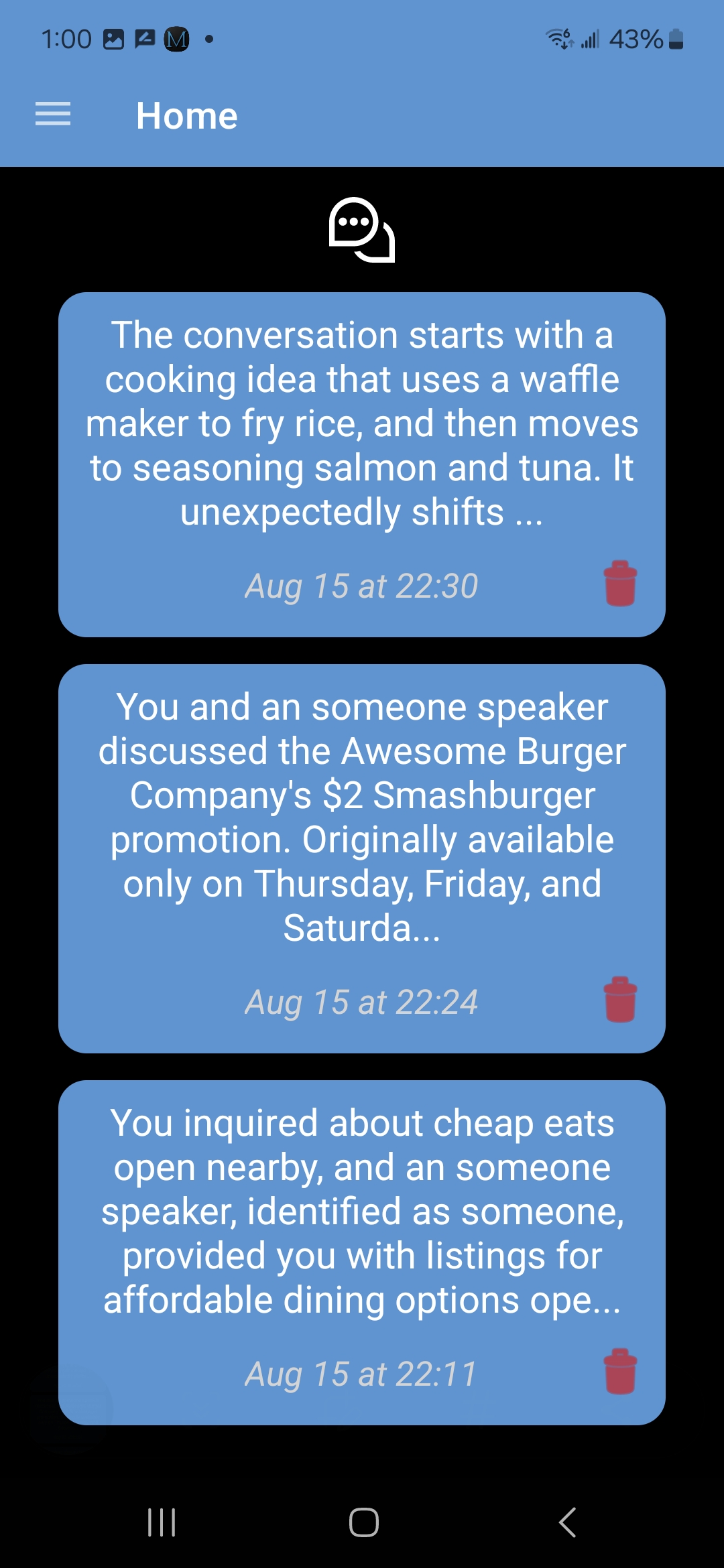}
    \caption{Conv. Dashboard}
    \label{fig:screenshot_conv_dashboard}
  \end{subfigure}
  \hfill
  \begin{subfigure}{0.19\linewidth}
    \includegraphics[width=\linewidth]{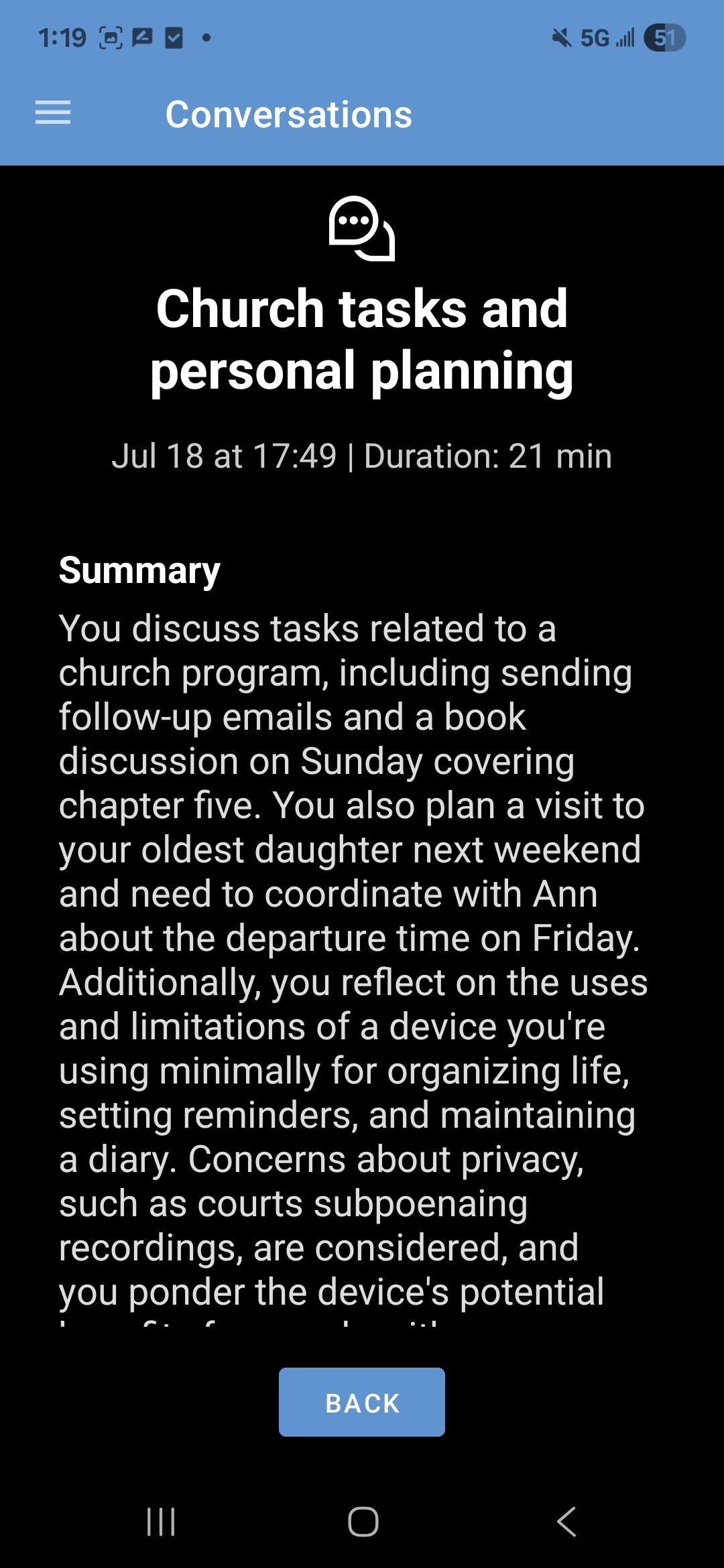}
    \caption{Conv. Summary}
    \label{fig:screenshot_conv_summary}
  \end{subfigure}
  \hfill
  \begin{subfigure}{0.19\linewidth}
    \includegraphics[width=\linewidth]{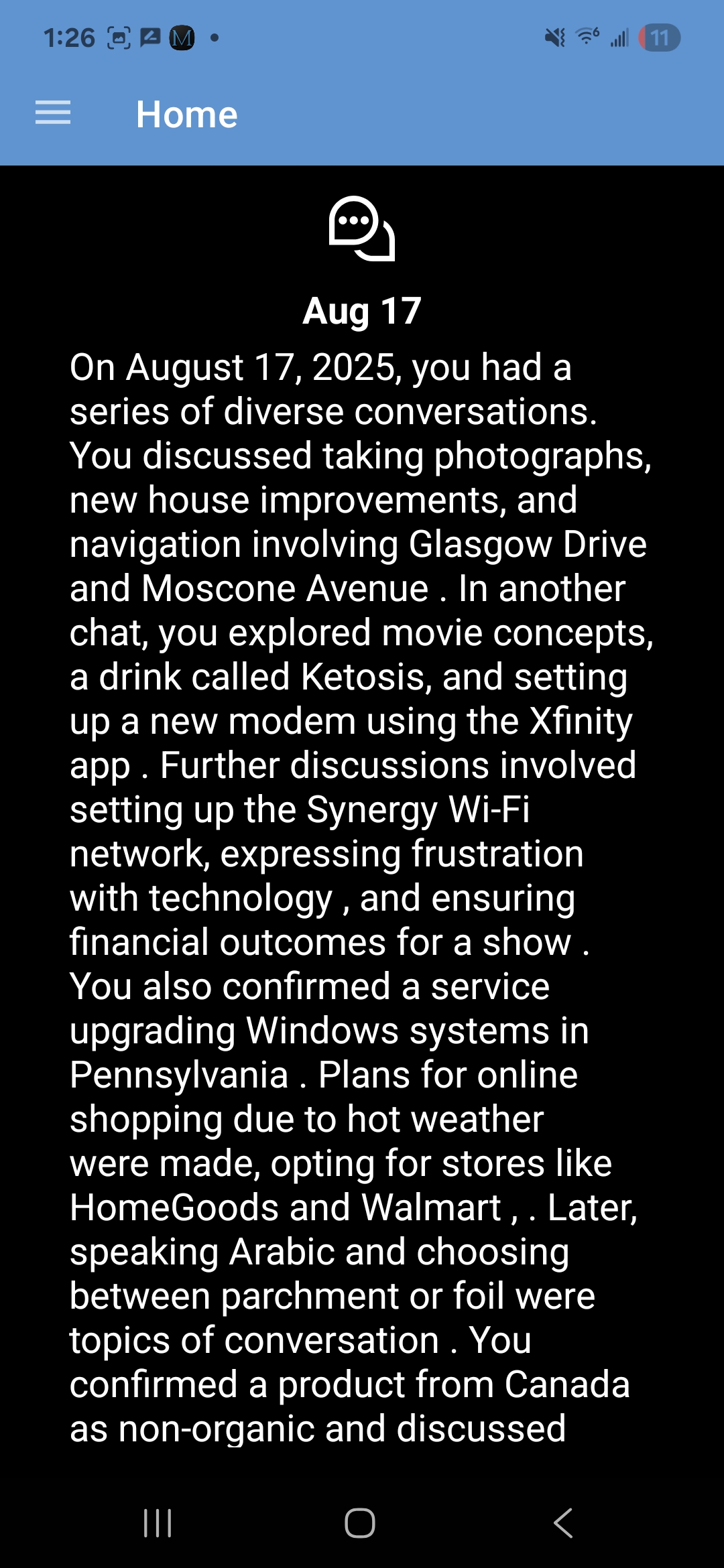}
    \caption{Daily Summary}
    \label{fig:screenshot_daily_summary}
  \end{subfigure}
  \hfill
  \begin{subfigure}{0.19\linewidth}
    \includegraphics[width=\linewidth]{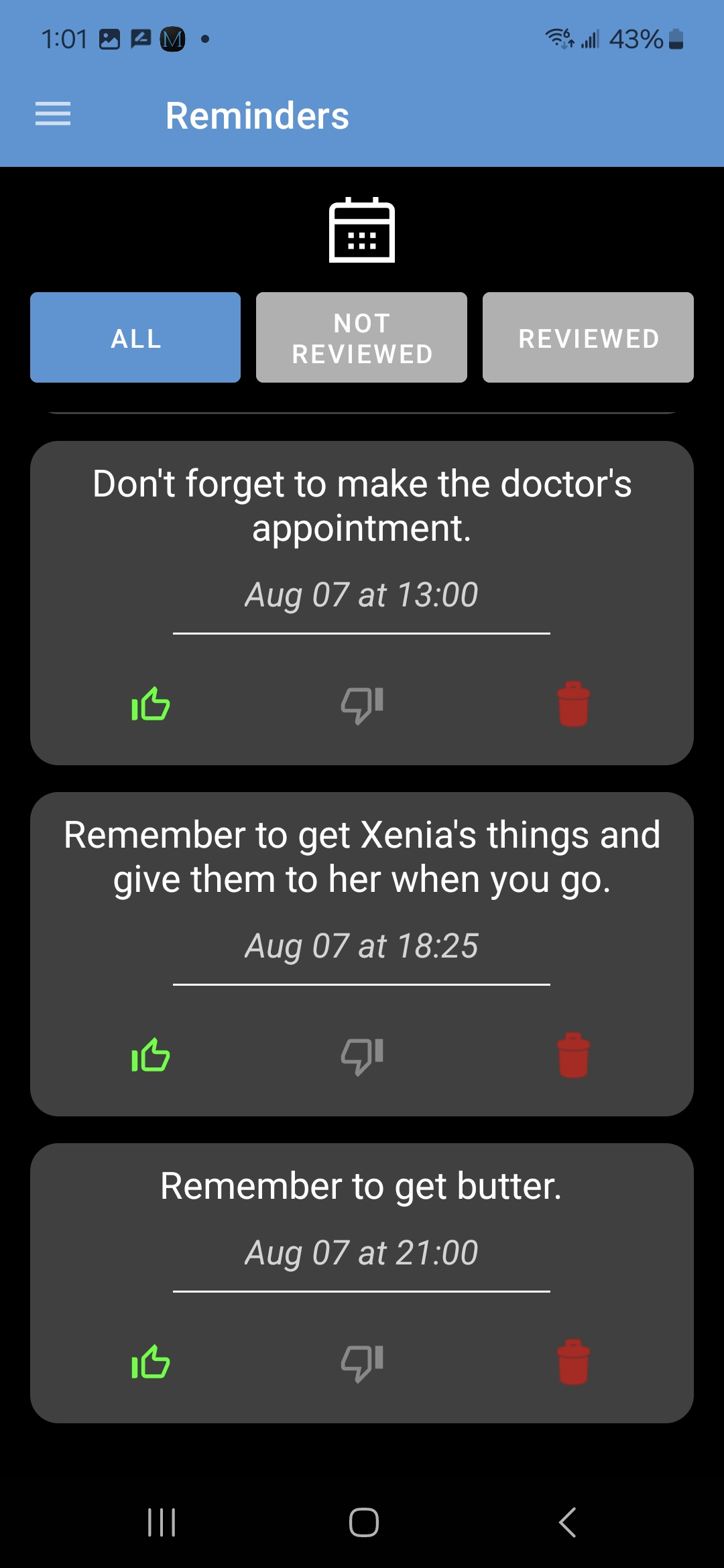}
    \caption{Reminders Dashboard}
    \label{fig:screenshot_reminders}
  \end{subfigure}
  \caption{Screenshots of the \textsc{Memoro 2} app interface showing the recording screen, conversation dashboard, conversation summary, daily summary, and reminders.}
  \Description{App UI Overview. Five screenshots illustrate the phone interface: home, recording screen, conversation dashboard, conversation summary, and reminders, depicting how users capture and retrieve memories.}
  \label{fig:screenshot_app_screens}
\end{figure*}

The mobile dashboard allows users to review/manage/delete memories. The dashboard supports three primary functions: (1) start/stop continuous audio recording; (2) view or delete past conversations, reminders, and recall queries (with responses); and (3) modify or delete scheduled reminders. Screenshots are shown in Fig. \ref{fig:screenshot_app_screens}.

\subsection{Retrospective Memory Augmentation}

To recall a detail from a previous conversation, the user speaks the trigger phrase with a contextual question (e.g., "Hey Navi, what time was that appointment again?"). To support recall, \textsc{Memoro 2} implements a context-aware AI agent using the GPT-4o Realtime API. The agent leverages session state to manage multi-turn interactions with the user, such as clarifying details they get through the system response or resolving conflicting information. The agent constructs a retrieval context by combining the 10 seconds of audio surrounding the query with the preceding 10 minutes of conversation. It also incorporates current-day and weekly summaries, enabling both date-specific and long-term queries.

\subsubsection{Retrieval Mechanism and Tools}
The agent has access to tools that identify memory chunks from past conversations:

\begin{itemize}
    \item \textbf{Retrieve More Memories}: Performs semantic search across stored chunks and returns the top 10 relevant chunks as summaries and their unique IDs. The ranking of chunks are using a weighted ranking formula for \textit{Recall Probability} ($P$), adapting the retrieval scoring of generative agents~\cite{park_generative_2023}:
\begin{equation}
\label{eq:recall_probability}
P = \alpha(\text{Recency}) + \beta(\text{Significance}) + \gamma(\text{Relevance})
\end{equation}
Where the weights were tuned during pilot iteration, values are in Appendix~\ref{appendix:meta:params}.
    \item \textbf{Get Ongoing Conversation}: Accesses the transcripts from the current conversation beyond the immediate 10-minute query window.
    \item \textbf{Get Reminders}: Queries the database of previously extracted commitments and social promises.
    \item \textbf{Get Full Conversation With Id}: Fetches original transcripts for more context when chunk summaries are insufficient for the query.
    \item \textbf{Manage Reminders}: Allows the agent to update or delete reminders if needed.
\end{itemize}

A safeguard limits the agent to five tool calls per query, triggering a standard fallback response if the target information cannot be found.

\subsubsection{Response Generation and Output}

The agent outputs structured JSON containing the response text and source timestamps to avoid attribution errors. When information is insufficient, it responds with "I can’t seem to find relevant memories" to counter hallucinations. Responses are optimized for text-to-speech delivery (using ElevenLabs API) via the headset, with natural phrasing in under 150 words \cite{ElevenLabs}.

\subsection{Prospective Memory Augmentation}
\label{subsec:prospective}

Prospective memory support here amounts to tracking commitments in conversation: detecting them in everyday speech, converting informal time references into concrete times, and updating them as they change across conversations. The system processes transcripts at the end of each conversation to detect actionable items from any speech heard, but only those relevant to the registered user. These commitments are then scheduled and delivered as audio prompts through the headset or as mobile notifications at the appropriate time. Reminders are updated by the agent as commitments evolve. For instance, coordinating a dinner might involve tracking who suggested it, when it was proposed, and the final agreed time.

\subsubsection{Agent Architecture and Processing Pipeline}

Similar to the recall system, the reminder agent establishes a WebSocket connection to OpenAI's GPT-4o realtime API with specialized tools for reminder management. The agent is equipped with two primary tools: \begin{itemize} 

\item \textbf{Retrieve More Memories}: Performs semantic search across stored conversations to gather additional context when commitment details are ambiguous or incomplete 

\item \textbf{Manage Reminders}: Creates, updates, or deletes user reminders based on confirmed commitments made during conversations, with each reminder assigned a unique UUID.

\end{itemize}

\subsubsection{Temporal Processing and Scheduling}
\label{subsubsec:temporal}

Informal time references (e.g., “tomorrow afternoon”) for the commitments are converted into absolute timestamps (MM-DD-YYYY, HH:MM:SS) using conversational context by the agent. Reminders are constrained to reasonable ranges, excluding 11 PM–6 AM and more than two weeks to keep it within the study duration.  For deadlines, a reminder is scheduled 30 minutes prior. Reminders are stored in an encrypted local vector database and scheduled with Android’s WorkManager for reliable delivery. The system maintains bidirectional sync with the device’s scheduler, supporting updates and cancellations when context or user behavior changes.

\section{Field Study Design}
We conducted a 10-day field deployment of \textsc{Memoro 2} as a technology probe to examine how older adults engage with an always-on wearable memory agent in everyday life. Ten days follow comparable free-living, in-situ studies of memory aids with older adults \cite{chan_prospecfit_2019}, balancing observation against participant burden. The study had no scripted tasks, allowing usage patterns and reflections to emerge naturally. The institutional review board approved the protocol. Participants received an explanation of the purpose, functions, risks, and data handling, and provided consent before participation. Each participant was given a study smartphone (Samsung Galaxy A35) running the \textsc{Memoro 2} app, a lightweight bone-conduction headset (Shokz OpenRun Pro 2), and an “I’m recording” badge that could optionally be used for bystander notification. The complete study workflow, from onboarding to the exit interview, is detailed in Fig. \ref{fig:memoro_study_workflow}. The study was approved by the university ethics review board.


\subsection{Participants}
Participants were recruited via two channels: email announcements to community members and in-person outreach at local senior centers. Eligibility required being age 60 or older; reporting no hearing or vision impairments that would interfere with using a wearable/voice interface; and having no formally diagnosed cognitive impairment. Because one of the study's goals was to co-design a future version of the technology, we also sought individuals who described themselves as curious or open to using everyday technology. Interested individuals completed a brief screening survey verifying these criteria; only those meeting all of them were invited to enroll. The final sample comprised 10 older adults (2 female, 8 male), $\mu_{age}=68.6, \sigma_{age}=6.6$). All participants spoke English as a first language, and 6 of them had used an AI assistant (Siri, Alexa, etc.) prior to the study. Participants received \$150 upon completion of the 10-day study; those who completed feedback in the app by rating the memory interactions received an additional \$50. Compensation reflected the substantial participant burden (lab visits, 10 days of system use, interviews every other day, and travel). There were no participant dropouts during the study. The distribution of the participants is available in Table \ref{tab:participants} and the recruitment survey in Appendix~\ref{appendix:meta:surveys}.
\begin{table}[t]
\centering
\small
\caption{Participant demographics for the field study. \textit{Confidence} is
self-reported confidence in using a new application; \textit{Prior AI}
indicates prior experience with an AI assistant.}
\label{tab:participants}
\begin{tabular}{@{}l c c l l c@{}}
\toprule
\textbf{ID} & \textbf{Age} & \textbf{Gender} & \textbf{Occupation} & \textbf{Confidence} & \textbf{Prior AI} \\
\midrule
P1  & 71 & M & Hotelier           & Extremely  & Yes \\
P2  & 86 & M & Retired            & Moderately & No  \\
P3  & 68 & M & Retired            & Extremely  & Yes \\
P4  & 68 & F & Retired            & Very       & No  \\
P5  & 70 & M & Production Manager & Extremely  & No  \\
P6  & 65 & M & Admin.\ Assistant  & Very       & Yes \\
P7  & 60 & F & Bookkeeper         & Very       & No  \\
P8  & 68 & M & Physician          & Extremely  & Yes \\
P9  & 66 & M & Engineer           & Very       & Yes \\
P10 & 64 & M & Engineer           & Extremely  & Yes \\
\bottomrule
\end{tabular}
\end{table}


\subsection{Onboarding Procedure}
\begin{figure*}[t]
  \centering
  \includegraphics[width=\linewidth]{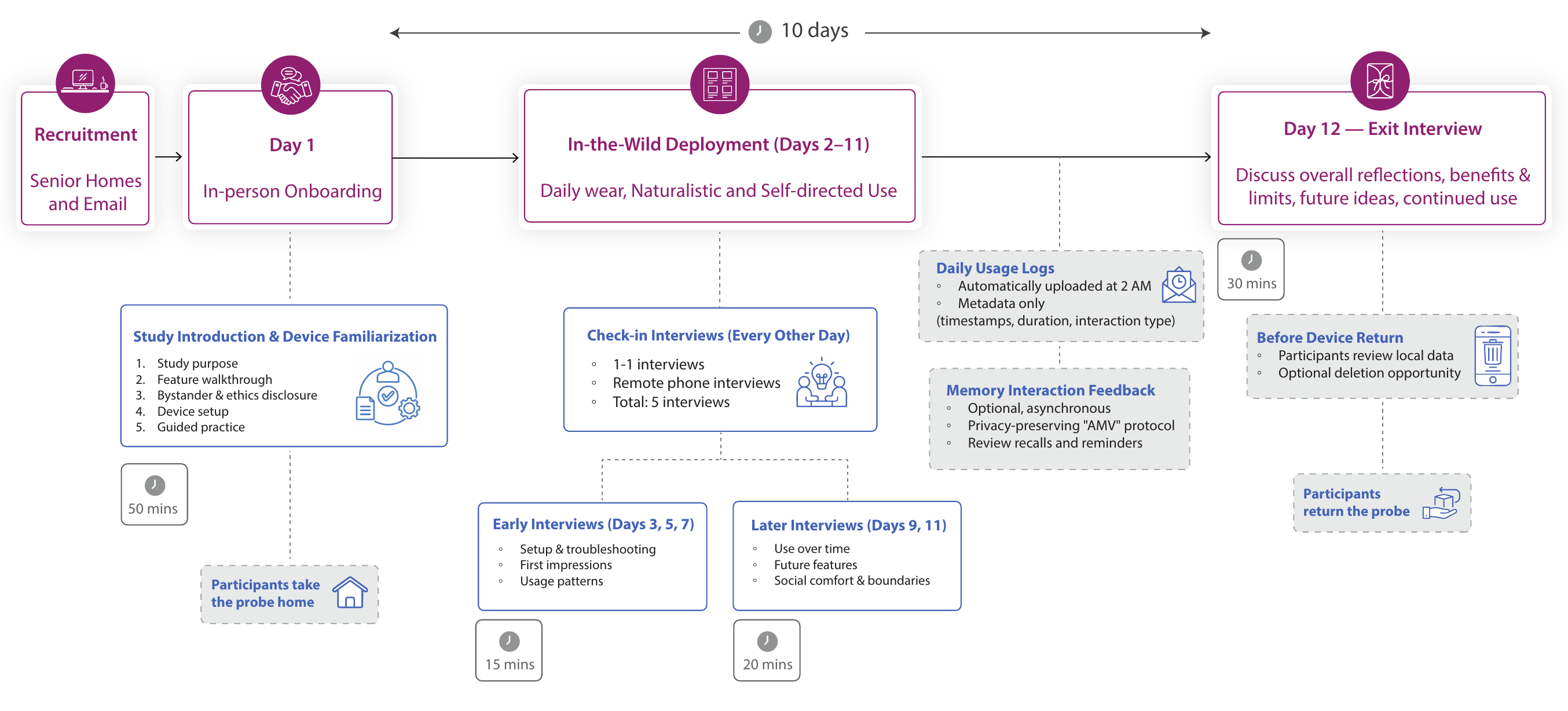}
  \caption{Study procedure: in-person onboarding (Day 1), a 10-day home deployment with check-in interviews every other day, nightly metadata uploads, and optional AMV feedback, followed by an exit interview with data review and device return (Day 12).}
  \Description{Flowchart outlining a 12-day study: recruitment, onboarding, 10-day field deployment with periodic interviews and logging, and a final exit interview including a data review and optional permanent deletion phase.}
  \label{fig:memoro_study_workflow}
\end{figure*}

Participants completed a 50-minute, on-campus lab onboarding session to familiarize themselves with \textsc{Memoro 2}’s features and operation before the study began, as shown in Fig. \ref{fig:memoro_study_workflow}.
\begin{enumerate}
\item Orientation \& Purpose (5 min): The experimenter introduced \textsc{Memoro 2} as a technology probe: the study aimed to understand real-world use of the device and gather suggestions for future designs. To ensure exposure across everyday contexts, we encouraged at least 4 hours of daily recording (actual recorded conversations may be lower, as transcriptions occur only when speech is detected in the recorded audio), with no penalties for not meeting the recommended time.

\item Feature Walkthrough (5 min): The experimenter showed videos of the core functions, such as the conversation summaries, recall, and reminders. Then they explained how the data was stored securely and locally, and described the methods for data deletion. 

\item Bystander Notification (5 min): As required by our Institutional Review Board (IRB), and in accordance with Massachusetts' all-party consent for audio recording \cite{MassWiretap}, participants were instructed to disclose that recording may occur in any setting where others are present, e.g., classrooms, work meetings, restaurants, public transit, and phone calls, and to pause use in spaces where recording is restricted. Alongside verbal disclosure, they could optionally wear a visible “I’m recording” button as an ambient cue. This practice follows prior work showing that explicit visual indicators increase bystander awareness of capture \cite{Koelle2018BeyondLED}. The experimenter walked the participant through a range of scenarios, such as someone entering the room during a group setting or an interruption mid-conversation, until the participant appeared comfortable using the system.

\item Hands-on Setup (5 min): Participants wore the bone-conduction headset and recorded their voice signature during setup.

\item Practice Time (30 min): Each participant held a semi-structured conversation with the experimenter (introductions and personal background). Using the designated trigger phrase ``hey navi,'' participants practiced querying conversation summaries and setting reminders.
\end{enumerate}


After onboarding, participants took the devices home, completed a 10-day home deployment with remote support via phone or text, and had no scheduled lab visits. Only those with recurring issues returned for in-person troubleshooting (P6, P7).

\subsection{Measures}
\subsubsection*{Interviews} We conducted brief 15-20 minute phone interviews every other day with each participant to gather feedback on use, barriers, and unique use cases. After the third interview, the questions moved to future situations they would like to use it for and the desired changes. A 30-minute exit interview focused on \textsc{Memoro 2} as a technology probe and opportunities for future designs. The phone interview questions and a sample of the final interview can be found in Appendix~\ref{appendix:meta:interviews}.

\subsubsection*{System Logs} The system logs captured anonymized usage metadata, including conversation metadata (start and end times, word and sentence counts), recall activity (timestamp, latency, type, user ID, and any user feedback), and reminder activity (creation and scheduled timestamps, type, user ID, and feedback if provided). For each interaction, the type was logged by prompting GPT-4o. The prompt for classifying the interaction into an assistance category is provided in the supplementary materials. Importantly, no conversational content was stored, and all transcripts were maintained solely on the device. 

\subsubsection*{Automated Memory Validation for System Evaluation} To evaluate the system while maintaining user privacy, we followed the Automated Memory Validation (AMV) protocol \cite{elagroudy_pixel_2025}. \citet{elagroudy_pixel_2025} designed the AMV as a method for in-situ evaluation where participants validate the system's output (e.g., accuracy of a reminder) directly on their own device, ensuring that researchers receive only the validation labels and optional free-text comments (e.g., "Helpful", "Inaccurate") without accessing the private content itself. In our system, participants rated each recall and reminder directly in the mobile application: thumbs-up for helpful/accurate responses, thumbs-down for inaccurate ones, with optional comments. Because we prioritized local-only storage, we could not access the ground truth data needed for traditional performance metrics (e.g., false-positive rates). Moreover, this was necessary for obtaining university ethics approval to run a continuous recording field study. AMV thus provides a privacy-preserving signal of perceived helpfulness on returned outputs. The offline component validation of \ref{subsec:offline_validation} complements this signal by testing whether the pipeline produces correct outputs against known ground truth.

\subsection{Offline Component Validation}
\label{subsec:offline_validation}

AMV rates only the outputs the system returned; it cannot count answers it never produced, commitments it missed, or content it fabricated. Participant data could not close this gap by design: the privacy architecture (Section~\ref{subsubsec:privacy}) keeps all conversational content on-device, precluding ground-truth annotation of participants' recordings. We therefore validated the recall and commitment-extraction components offline against ground-truth transcripts, feeding them through the deployed pipeline (production segmentation, embeddings, ranking, prompts, tools, and tool-call limits; parameters in Appendix~\ref{appendix:meta:params}). Transcription, diarization, and TTS use commercial APIs whose performance is characterized elsewhere; summarization was assessed through participants' in-situ use and interview feedback.

\subsubsection*{Validation Corpus}
A corpus of 14 conversations, sized to approximate one week of use based on estimates that older adults engage in roughly two social interactions per day \cite{luo2022more} was used to validate the system. The conversations  were constructed matching the kind of speech the system could encounter in deployment: everyday spoken dialogue of an older adult with family, friends, and service staff. The corpus seeds 45 ground-truth commitments spanning explicit times, informal time references, and details that change across conversations (Section~\ref{subsec:prospective}), plus 15 distractors (hypotheticals, other people's obligations, completed events) so that extraction precision is meaningful. Fifty recall questions matched how participants could use the probe: first-person spoken queries asked zero to ten days after the source memory, evaluated against the memory store as it existed at that moment, including 15 about content never discussed, testing whether the system declines rather than fabricates. This corpus is a validation instrument matched to our deployment, not a proposed benchmark. Answers were scored by an LLM judge against gold labels~\cite{zheng2023judging}; three blinded, independent human raters re-judged every recall answer and a stratified 19\% sample of reminder matches; the majority agreed with the judge on 96\% of questions and never rated as factually wrong an answer it passed (Appendix~\ref{appendix:meta:offline}).

\begin{table*}[t]
\centering
\caption{Offline component validation on the deployed pipeline.}
\label{tab:offline_validation}
\begin{tabular}{l p{6.5cm} l}
\hline
\textbf{Component} & \textbf{Metric} & \textbf{Result} \\
\hline
Recall QA & Accuracy: single-hop / multi-hop / temporal & 60\% / 80\% / 60\% \\
Recall QA & Answer rate; accuracy when answering (single / multi / temporal) & 91\%; 72\% (64\% / 89\% / 67\%) \\
Abstention & Unanswerable questions correctly refused & 14/15 \\
Commitment extraction & Precision / Recall / F1 & 0.56 / 0.98 / 0.72 \\
Temporal grounding & Scheduled time exact; within acceptable window & 45\%; 89\% \\
Evolving commitments & Final state correct after cross-conversation updates & 3/5 \\
\hline
\end{tabular}
\end{table*}

Table~\ref{tab:offline_validation} summarizes the results. Across all 50 questions, counting correct refusals as correct, overall accuracy was 74\% (37 of 50). For a memory aid deployed with older adults, safety is critical: the system refused 14 of 15 unanswerable questions, captured 44 of 45 commitments, and scheduled 88.6\% of reminders within their acceptable windows. Of the 12 imperfect answers, six were correct but incomplete (e.g., a date without its time), one overly narrow, three declined, and two malformed with no answer. Commitment extraction errs on the side of extracting too much: one distractor of fifteen leaked through, and low precision comes mainly from duplicates rather than invented commitments; counting duplicates once raises precision to 0.79, pointing to cross-session deduplication as a design gap. Exact-time agreement trails window agreement because the agent extends the 30-minutes-early deadline rule (Section~\ref{subsubsec:temporal}) to informal time references. A technology probe must work in real-world use before it can surface user needs and design ideas~\cite{hutchinson_technology_2003}; this validation tests that requirement rather than benchmark performance. \textsc{Memoro 2} met it: the system refused what it could not answer and erred mostly by omission, so the findings in Section \ref{sec:results} reflect participants' experience with a functioning memory aid.

\subsection{Data Analysis}
Interviews were recorded, transcribed, anonymized, and securely stored in accordance with IRB protocol. We used  thematic analysis \cite{braun_clarke_2022_thematic} with a hybrid inductive--deductive approach. Across 60 interviews and system logs, we followed a four-stage iterative process: (1) familiarization with transcripts and recordings, linking excerpts to research questions; (2) integration of qualitative data with participant-level usage summaries from logs; (3) iterative coding refinement through independent coding and negotiated agreement \cite{oconnor2020}; and (4) theme development and validation against the full dataset. Quantitative log data are reported descriptively and integrated throughout. \textbf{\textit{Codebook in Supplementary Materials.}}

\subsubsection*{Offboarding Data} At the end of the study, participants were offered the opportunity to review and delete any and all data (conversations, recall, and reminders) and, optionally, share selected data with the research team.

\section{Results}
\label{sec:results}

Our findings draw on both system usage logs and semi-structured interviews to characterize how older adults engaged with the technology probe over the 10-day study period. We organize the results around our research questions, integrating usage patterns with participants' reflections.

\subsection{Context-Dependent Adoption: Daily Rhythms and Drivers of Engagement}
\label{subsec:rq1}

\begin{figure*}[h]
  \centering
  \begin{subfigure}{0.36\linewidth}
    \centering
    \includegraphics[width=\linewidth]{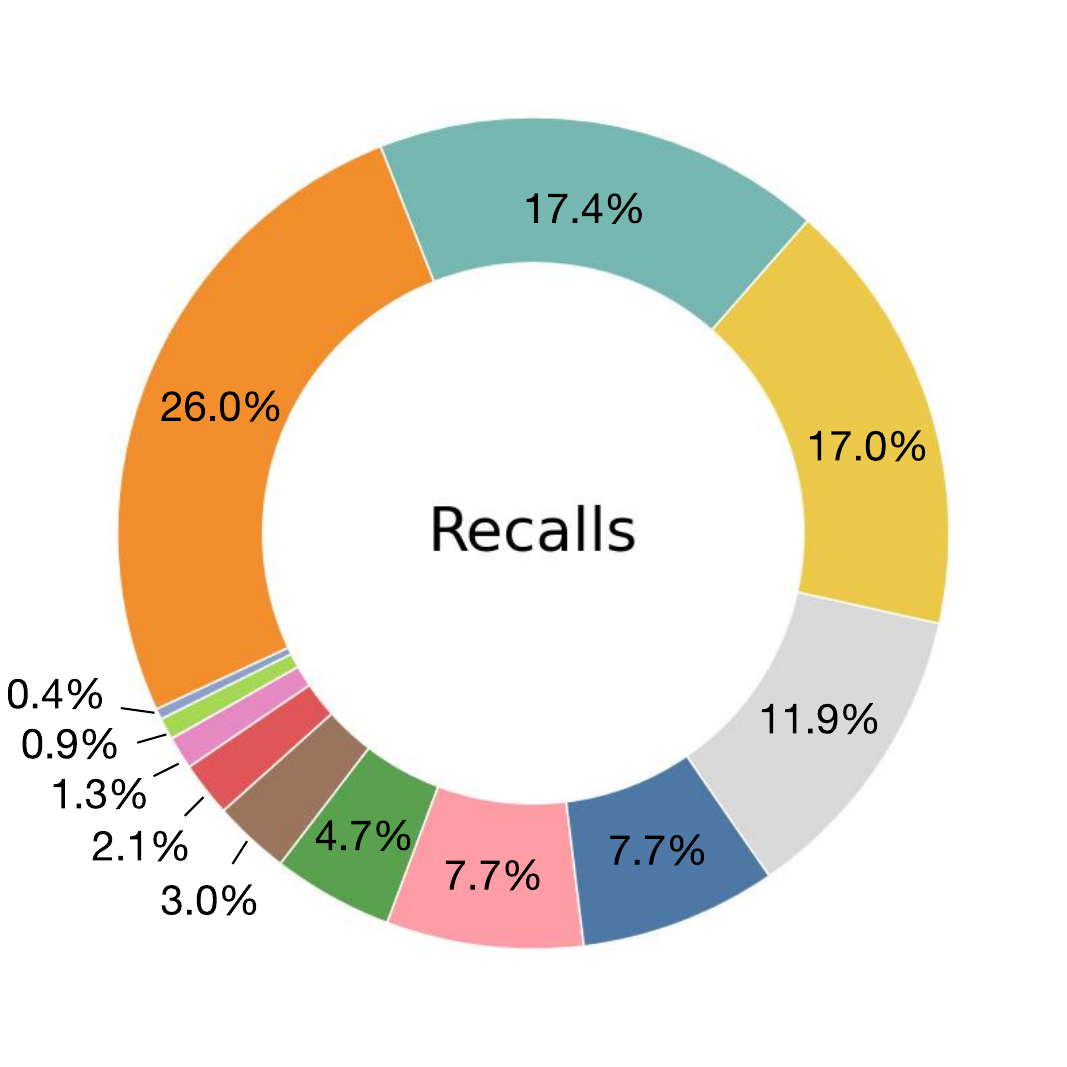}
    \caption{Recall types distribution}
    \label{fig:recall_pie}
  \end{subfigure}\hfill
  \begin{subfigure}{0.36\linewidth}
    \centering
    \includegraphics[width=\linewidth]{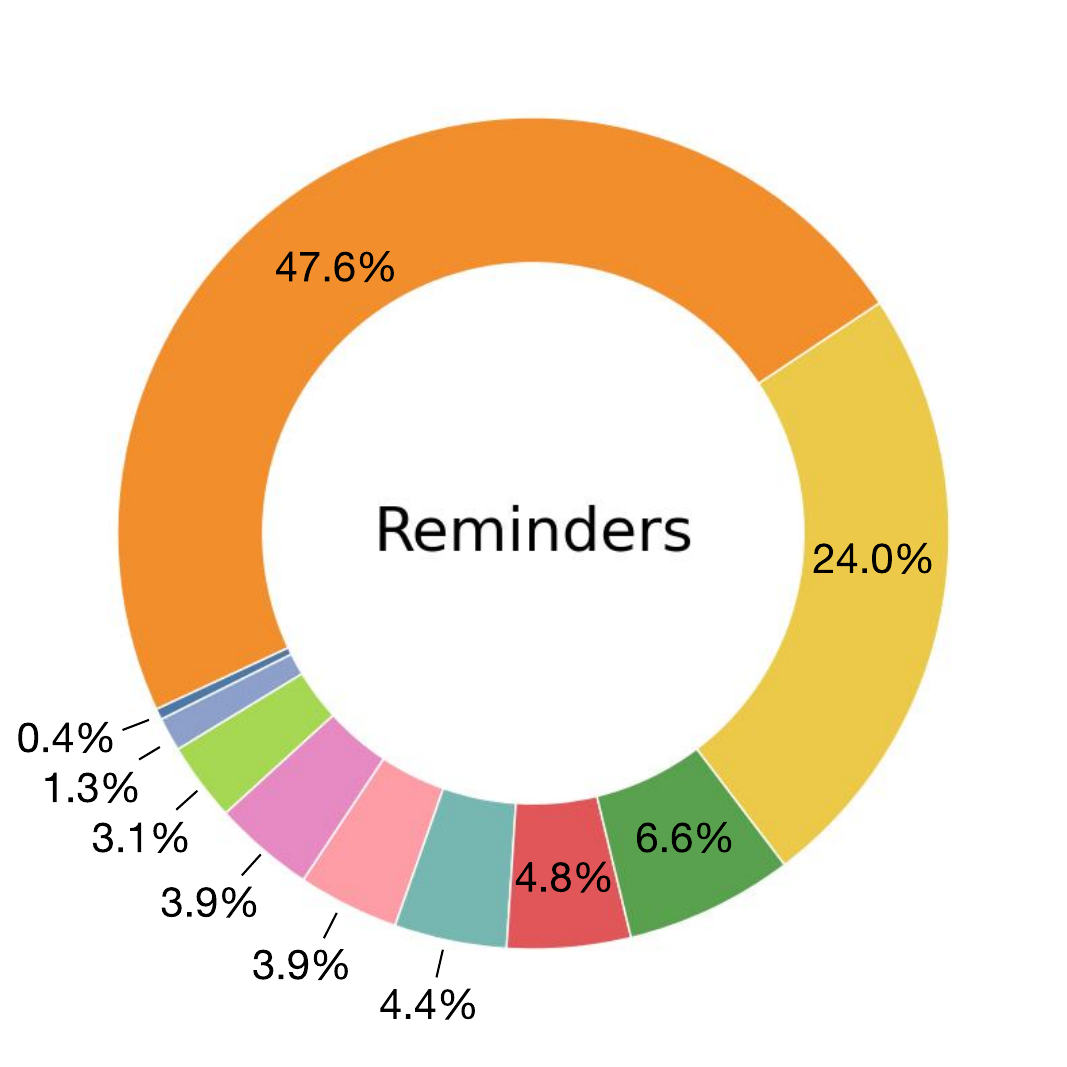}
    \caption{Reminder types distribution}
    \label{fig:reminder_pie}
  \end{subfigure}\hfill
  \begin{subfigure}{0.24\linewidth}
    \centering
    \raisebox{.5\height}{\includegraphics[width=\linewidth]{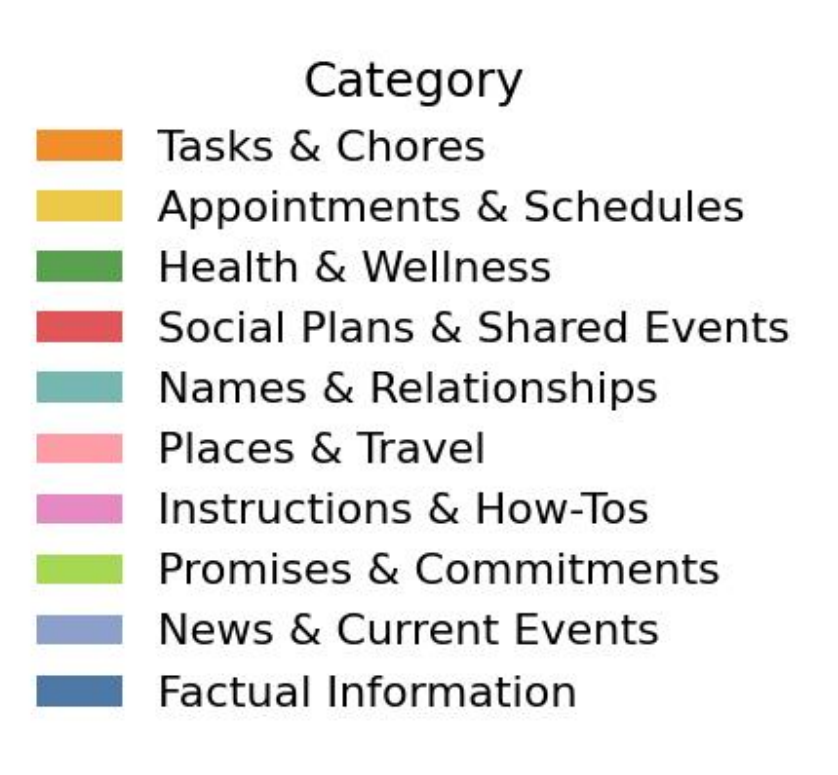}}
    \label{fig:legend}
  \end{subfigure}

  \vspace{1em}

  \begin{subfigure}{0.9\linewidth}
    \centering
    \includegraphics[width=\linewidth]{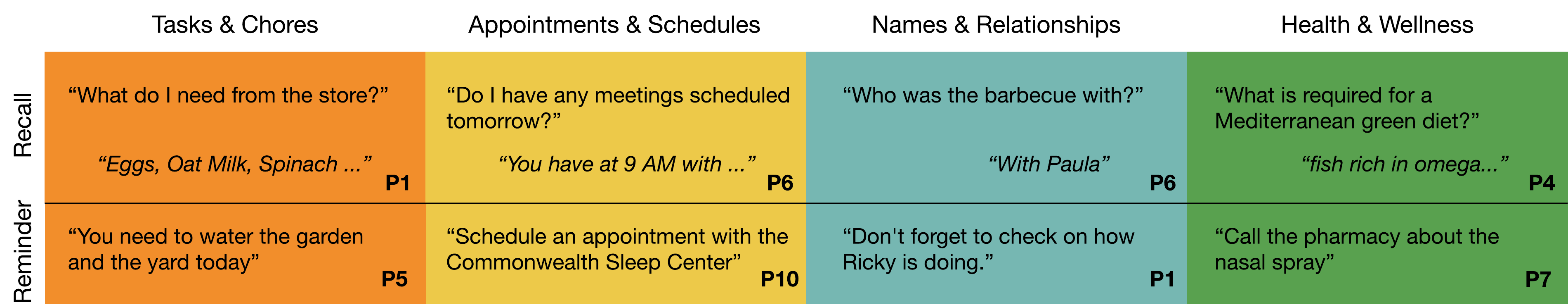}
    \caption{Examples of system interactions during the 10-day field study}
    \label{fig:examples}
  \end{subfigure}

  \caption{(Top) Distributions of recall and reminder types used by participants during the 10-day field study. (Bottom) Examples of system interactions.}
  \Description{Recall vs. reminder categories. Donut charts show category distributions: recalls and reminders are dominated by tasks/chores and appointments; a panel of example queries illustrates typical interactions across categories.}
  \label{fig:recall_reminder_examples}
\end{figure*}

To explore how older adults adopt an always-on memory agent, we first examine the empirical usage patterns from the system logs. Across both recalls and reminders, everyday task management emerged as the primary use case: Tasks and Chores accounted for the largest share of both recalls (26\%) and reminders (47.6\%). However, while recalls also frequently supported social memory through Names and Relationships (17.4\%), reminders were more strongly oriented toward future activity planning, with Appointments and Schedules comprising a quarter of all reminder requests (24\%). Category rankings and example interactions are shown in Fig. \ref{fig:recall_reminder_examples}.

\subsubsection{Usage Patterns: Engagement and Daily Rhythms}

\begin{figure*}[h]
  \centering
  \begin{subfigure}{0.49\linewidth}
    \includegraphics[width=\linewidth]{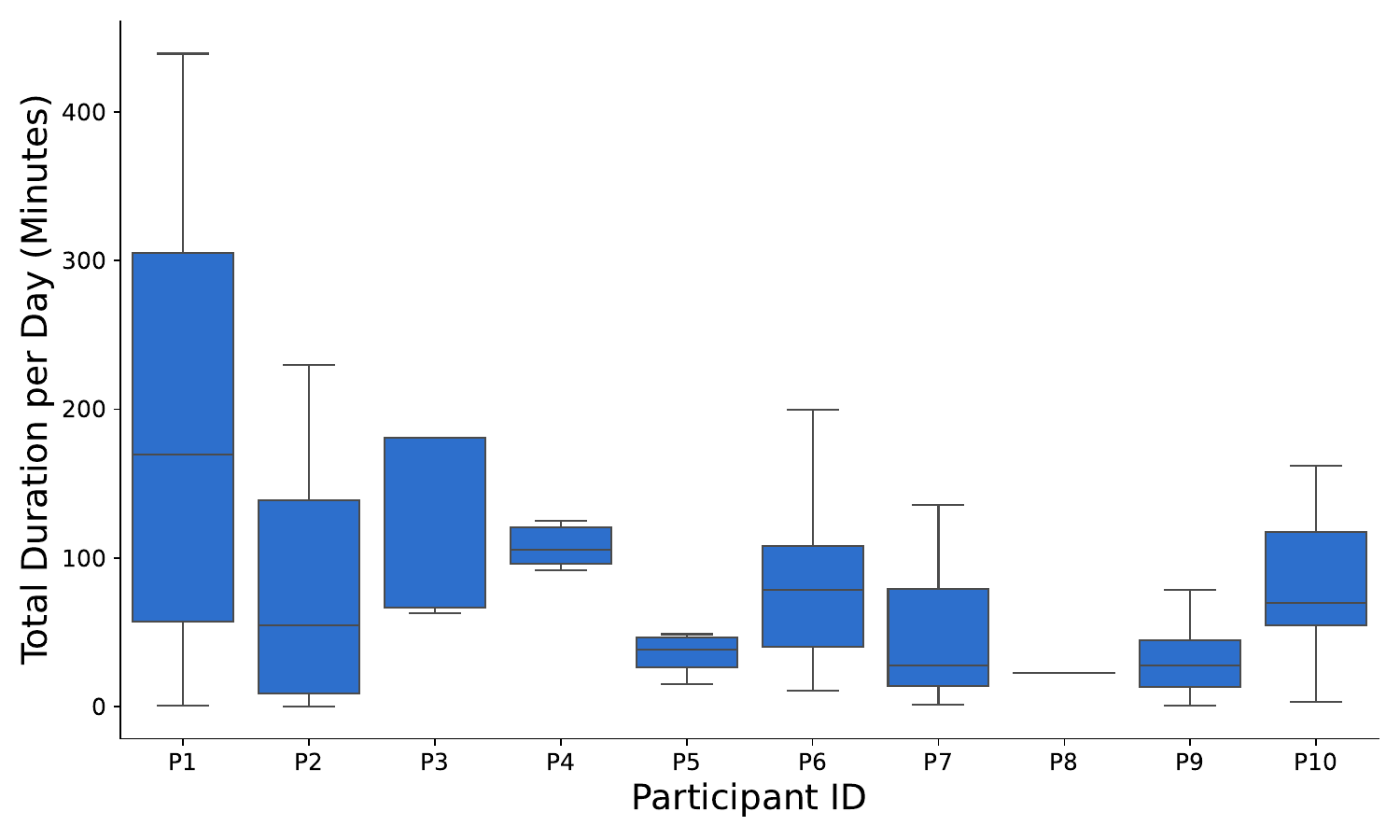}
    \caption{Total duration of conversations per participant per day since start of the study, showing varied usage between participants}
    \label{fig:user_avg_duration}
  \end{subfigure}
  \hfill
  \begin{subfigure}{0.49\linewidth}
    \includegraphics[width=\linewidth]{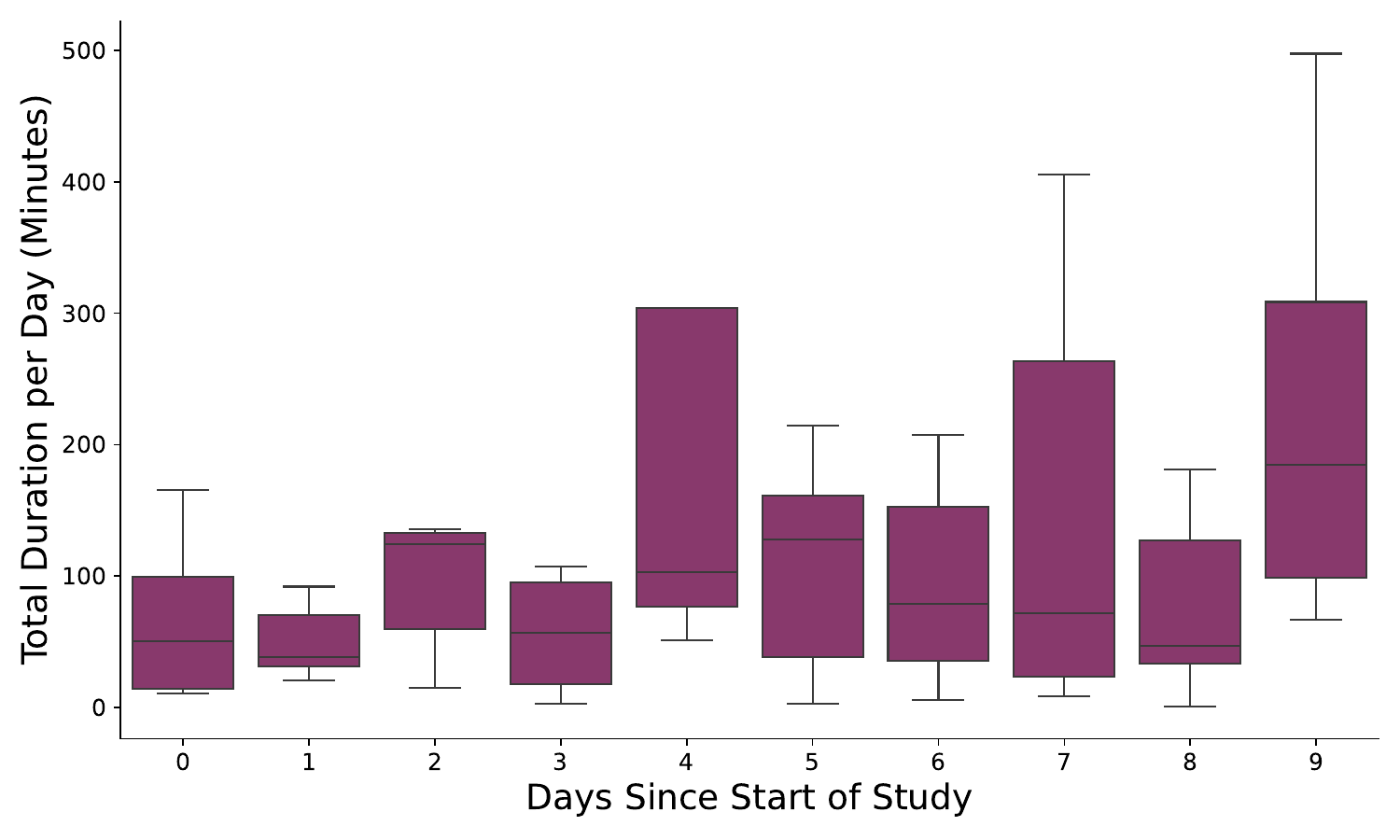}
    \caption{Total duration of conversations per day since start of the study, showing consistent usage over the study period}
    \label{fig:avg_usage_by_day}
  \end{subfigure}
  \caption{Distributions of duration of captured conversations during the 10-day field study by the participants using the always-on audio agent}
  \Description{Figures showing the distribution of \textsc{Memoro 2} using a boxplot. Boxplots summarize (a) per-participant daily conversation duration and (b) duration by study day; they reveal large individual differences, several heavy users, and generally steady use over the 10 days rather than a systematic drop-off.}
  \label{fig:user_conversation_durations}
\end{figure*}

Participants integrated the system into their daily lives in different ways. Usage variation reflected differences in lifestyle and situational context rather than disengagement, as participants noted that social settings, personal routines, and confidence in the system's functionality influenced when and how they used \textsc{Memoro 2} (see Fig. ~\ref{fig:user_conversation_durations}). Variation across participants and across days aligns with findings from EAR-based (electronically activated recorder) research on older adults' everyday conversation, where daily speech engagement varies substantially between individuals (driven by partnership status, household composition, and routines) and across days within a single person~\cite{demiray2020sounds, mehl2017ear}. Despite this variation, overall engagement remained stable across the 10-day study period, with usage characterized by intermittent bursts of activity rather than a systematic decline over time.

\begin{table*}[h]
\caption{Summary of System Usage and Engagement}
\label{tab:usage}
\centering
\begin{tabular}{p{6cm} p{5cm}}
\toprule
\textbf{Metric} & \textbf{Value} \\
\midrule
Total in-conversation speech transcribed & 144.2 hours \\
Mean audio recorded per participant & 9.8 hours ($\sigma$=14.8) \\
Total conversations & 939 \\
Mean conversation length & 9.2 minutes ($\sigma$=36.2) \\
Total interactions & 464 \\
User-initiated recalls & 235 \\
Proactive reminders & 229 \\
Recalls rated helpful & 96\% (209 / 217 rated) \\
Reminders rated helpful & 92\% (136 / 148 rated) \\
\bottomrule
\end{tabular}
\end{table*}

\begin{figure*}[htbp]
  \centering
    \includegraphics[width=0.7\linewidth]{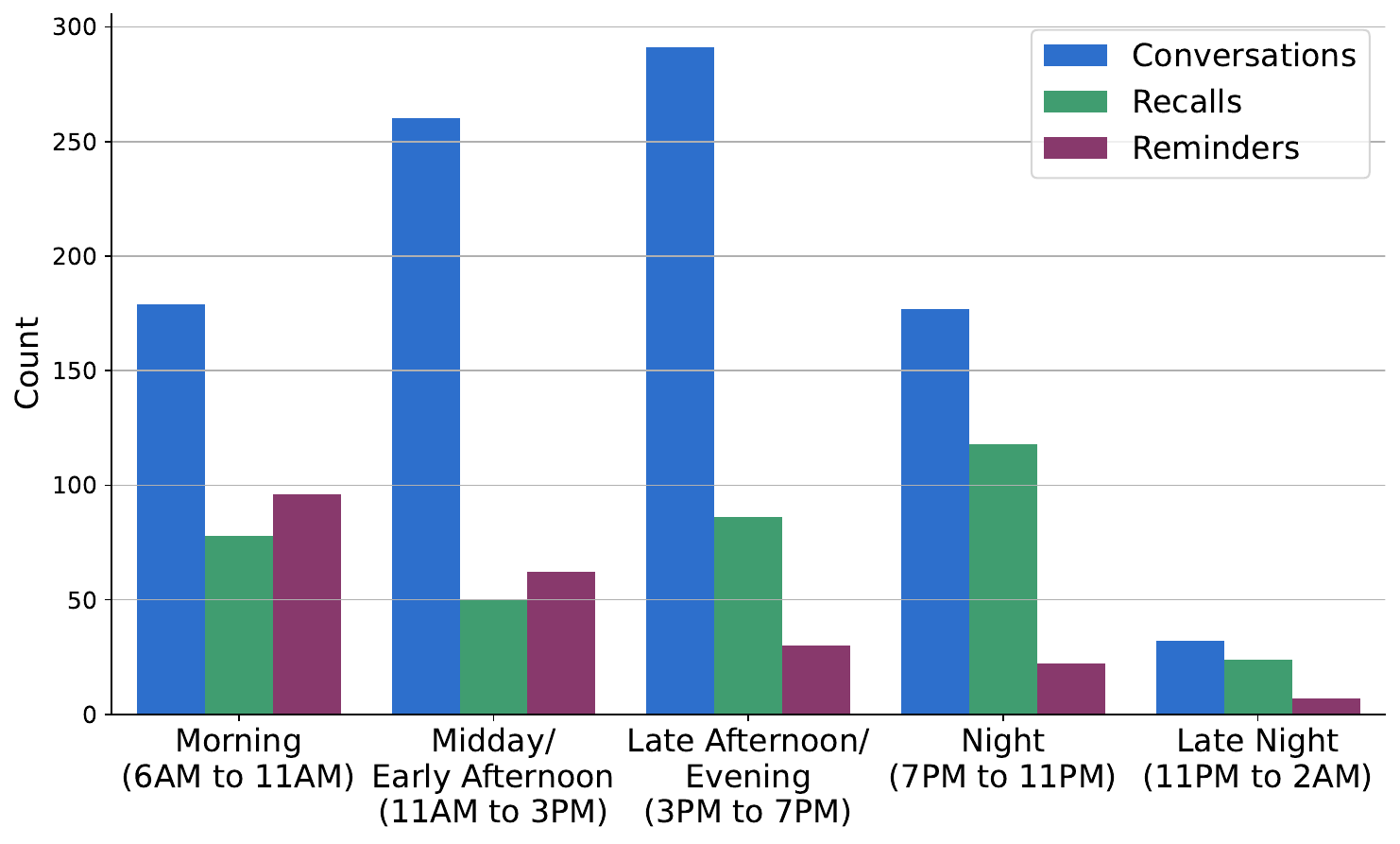}
    \caption{Distribution of conversation capture, recalls, and reminders throughout a 24-hour day during the field study}
  
  \Description{Time-of-day patterns. Bar chart of counts by time window shows conversations most frequent from midday through evening, recalls peaking later in the day and at night, and reminders concentrated in the morning and tapering toward night.}
  \label{fig:time_of_day}
\end{figure*}

Participants used the agent's two modes in near-equal measure (235 user-initiated recalls, 229 proactive reminders) but at different times of day. Across participants, interactions followed daily rhythms. Conversations most frequently occurred from late morning through early evening, reminders were concentrated in the mornings to cue daily actions and support planning (6--11 AM), while recalls peaked at night (7--11 PM), suggesting it as a form of end-of-day review (see Figure~\ref{fig:time_of_day}). Together, these patterns indicate that the system was incorporated into existing routines aligned with participants’ daily temporal structures.

Quantitative summaries of system usage and engagement are provided in Table~\ref{tab:usage}. Notably, the reported duration of conversational audio reflects speech segments retained after Voice Activity Detection (VAD) filtering, which removes non-speech audio from the recording. Overall perception was strongly positive: of the 365 interactions rated (by 8 of 10 participants), 94\% were positive. More details in Table \ref{tab:usage}). Negative feedback primarily reflected breakdowns in conversational turn classification or mismatches in reminder timing.


\subsubsection{Drivers of Continued Use: Routine Support and Ambient Documentation}


Participants described continued integration as being driven by two value propositions:

\begin{enumerate}
    \item Supporting daily routines with recalls and reminders
    \item Synthesizing the day through ambient documentation
\end{enumerate}

\begin{figure*}[h]
  \centering
    \includegraphics[width=\linewidth]{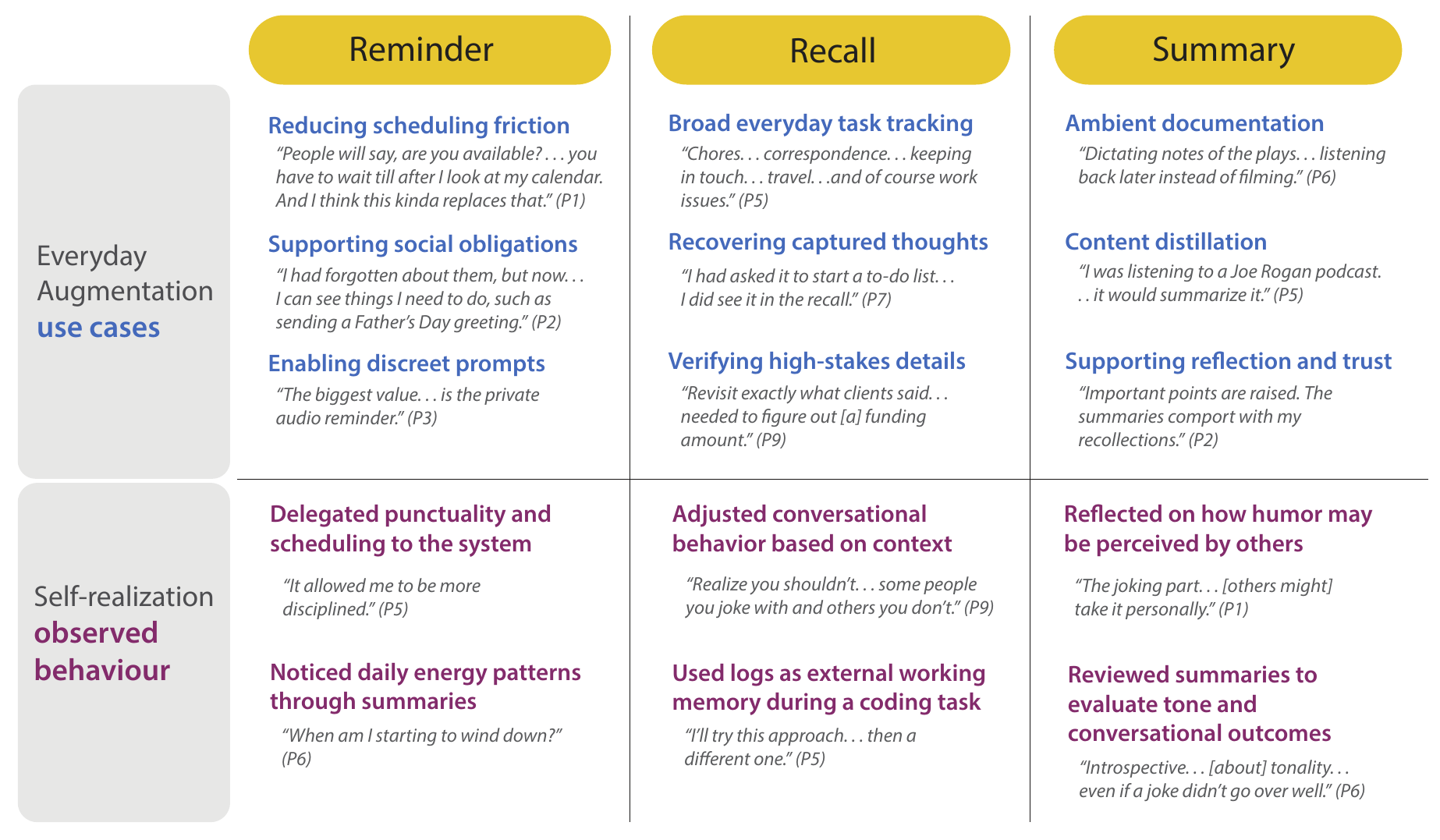}
    \caption{Use cases and observed behavior related to \textit{Self-Reflection and Planning Daily Tasks}, organized by behavioral categories as rows and the system features as columns}
    \Description{A two-part table organized by three system features: Reminder, Recall, and Summary, as column headers. The upper section, Self-Reflection and Planning Daily Tasks, contains three rows of participant quotes per feature, each preceded by a bold insight label. The lower section, Social Dynamics and Performativity while Recording, contains two rows of participant quotes per feature, also preceded by bold insight labels. Row labels on the left identify the two sections. All quotes are attributed to participants P1 through P9.}
    \label{fig:use-cases-and-observed-behaviors}
  \hfill
\end{figure*}

Representative quotes are summarized in Fig~\ref{fig:use-cases-and-observed-behaviors}. These patterns were consistently expressed across participants and usage contexts.

\paragraph{Supporting daily routines with recalls and reminders.}
A major driver of adoption was the system’s ability to support daily routines around remembering and acting on information. Some emphasized the role of reminders in prompting action without extra effort, with P7 reporting: “I had asked it to start a to-do list… I did see it in the recall.”. A few highlighted how recalls reinforced memory, as P2 noted: “Did find a conversation summary... to be very lucid, brief, and to have hit the high points.” These functions often replace or augment more effortful practices like manual note-taking or using a calendar.

In higher-stakes contexts such as conversations with a lawyer about a case, therapy sessions, medical appointments, or legal/financial consultations, participants emphasized the value of accuracy and traceability. Compared to manual note-taking or dictation, system-generated synthesis was easier to integrate and more effective in supporting follow-through. Some used recalls to revisit exact wording from prior conversations, as P9 noted: “Revisit exactly what clients said… needed to figure out [a] funding amount”.  Others highlighted the system’s role in easing social or situational friction. Rather than interrupting conversations to consult a phone or calendar, participants could rely on the system to surface relevant information later, supporting more fluid interactions in the moment.

Reminders were especially appreciated when they could be delivered discreetly. Participants noted that private audio reminders allowed them to receive prompts without drawing social attention or disrupting ongoing activities, and one participant described it as "the biggest value" they get from \textsc{Memoro 2} (P3). This sense of subtle, always-on assistance contributed to perceptions of usefulness and ease of use, aligning with established models of technology acceptance \cite{Peng2022Wearable}. 

\paragraph{Synthesizing the day through ambient documentation.}
Participants valued end-of-day summaries as concise textual reviews that replaced the need to reread full transcripts. These summaries were used to reconstruct daily events, revisit conversations, and recall media consumed earlier in the day. One participant used the summaries to record his son's football plays and used \textsc{Memoro 2} to dictate "notes of the plays" (P6). Described as trustworthy and "comport with my recollections"(P2), they positioned \textsc{Memoro 2} not just as a comprehensive memory archive but also as a tool for reliable synthesis.

\subsection{Behavioral and Social Adaptation Through Always-On Presence}
\label{subsec:rq2}


We examined how the continuous presence of \textsc{Memoro 2} shaped participants' everyday behaviors and social interactions. Participants reported that the system influenced not only what they remembered, but how they reflected, planned, and interacted with others.

\subsubsection{Reflection on Conversational Behavior and Social Planning} 


Participants reported that reviewing summaries and recalls increased awareness of their own communication and daily conversational behavior. Several described noticing tone, humor, and conversational context after revisiting conversation and daily summaries, including cases where humor misfired: “Introspective… [about] tonality… even if a joke didn’t go over well” (P6), and “The joking part… [others might] take it personally” (P1). The system was used to support social planning by capturing conversations and generating commitments and follow-up actions. Participants used reminders to stay “more disciplined” and relied on conversation logs to track and compare different approaches as they solved a problem (P5).  In professional or task-focused contexts, participants incorporated conversation logs and summaries into their workflow to observe daily rhythms such as changes in energy or focus. Representative quotes from participant interviews are shown in Fig. \ref{fig:use-cases-and-observed-behaviors}.

\subsubsection{Social Dynamics and Performativity while Recording}

\begin{figure*}[h]
  \centering
    \includegraphics[width=\linewidth]{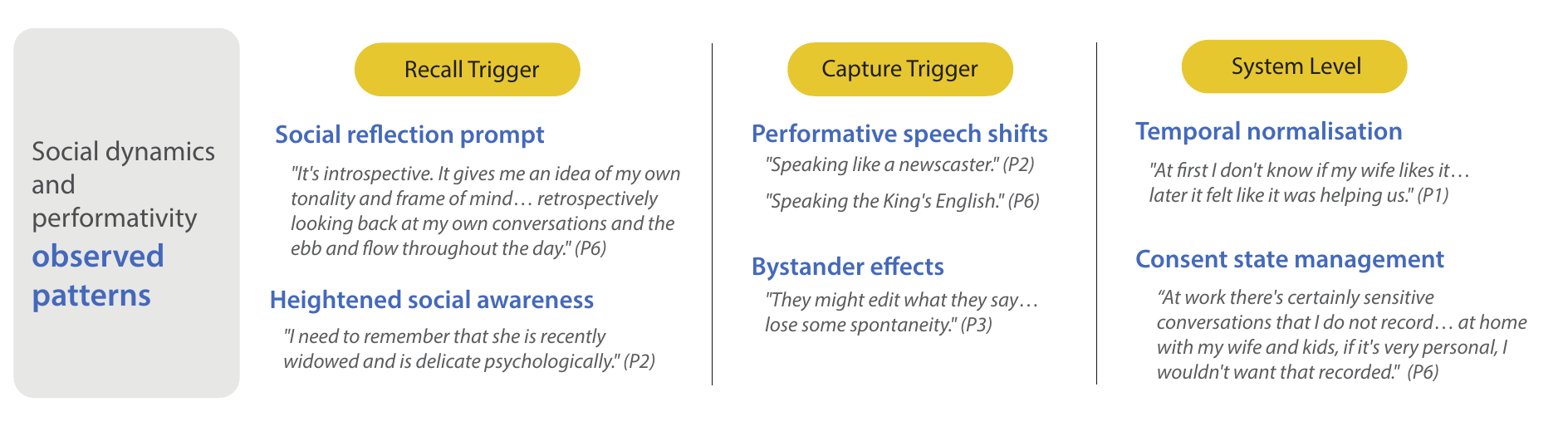}
    \caption{Observed patterns related to \textit{Social Dynamics and Performativity while Recording}, organized into three categories: Recall Trigger, Capture Trigger, and System Level.}
    \label{fig:social-dynamics}
  \hfill
\end{figure*}

The always-on presence of \textsc{Memoro 2} influenced social interactions such as conversations, as shown in Fig. \ref{fig:social-dynamics}. Some participants described initial discomfort or tension in close relationships related to recording conversations, followed by later use of the system to revisit prior discussions or clarify details. As one participant reflected early in the study, “At first I don’t know if my wife likes it…” and towards the end of the study emphasized that “it felt like it was helping us,” suggesting a move from uncertainty toward shared, normalized use (P1). Summaries were used to recall contextual details relevant to sensitive interactions. Additionally, participants noted changes in speech when they were aware of being recorded, including speaking more formally and in the "King's English" as one participant described (P6) during early use. Some expressed concern that recording might cause conversation partners to self-edit, though several noted that this effect decreased over time.

\subsection{Participants Sought Discreet Control, Adjustable Feedback, and Fit with Existing Tools}
\label{subsec:rq3}

Participants’ lived experience with \textsc{Memoro 2} surfaced the controls and fit participants sought from memory support wearables, ranging from integration with existing tools to low-friction interaction to transparency, curation, and privacy.

\subsubsection{Discreet and Low-Friction Interaction}

Voice interaction appealed to some participants for its hands-free control, but social context shaped which other interaction forms they wanted. They preferred simple voice commands for capture and control (e.g., record/stop/delete), but worried about using voice in group settings where it may feel awkward or draw attention. Participants imagined discreet alternatives such as small physical controls, subtle gestures, or other low-visibility interactions, so they could use the system without announcing it in such contexts. Importantly, participants also raised concerns about complicated consent management when additional people enter a conversation unexpectedly. Their experience suggested that capture is socially negotiated: a system that can be activated easily must also help users avoid accidentally violating expectations. In other words, low friction cannot only mean easy to start, but also easy to use responsibly.

\subsubsection{Transparency, Tunable Feedback, and Lightweight Curation}


Participants emphasized the importance of understanding what \textsc{Memoro 2} was doing at any moment. Confirmation that commands were received, and indicators for recording status, battery, or error states, were critical for trust. Yet, feedback needed to be adjustable: too much could be disruptive. One participant (P6) suggested a “temperature dial” to control the system’s presence—quiet in social spaces, more explicit when reliability mattered. These transparency needs intersect with preferences for selective capture and lightweight curation. Participants expressed that recording \emph{everything} was overwhelming: they wanted to prioritize meaningful moments and actionable information. Ideas included rolling buffers to preserve important segments and organizing content by type or time window to support quick review and action. Representative quotes from interviews and additional design suggestions are detailed in the supplementary materials.

\subsubsection{Fitting with Existing Tools and Routines}
A recurring ask was better integration with existing memory aids while adding another memory support system. P1 said they are “reliant on Microsoft Outlook and the calendar,” signaling that \textsc{Memoro 2} should plug into what already exists rather than replace it. P3 pushed for continuity at key moments, “A few minutes before a scheduled call, [Memoro 2] would say the event name and start time”, while P4, a longtime Dayminder(daily planner) user, asked for a “simple, glanceable layout” that mirrors familiar paper views. Together, these suggestions underscore a desire to extend current workflows and habits, not rebuild them.




\section{Discussion}
\label{sec:discussion}

\begin{figure*}[t]
    \centering
    \includegraphics[width=\linewidth]{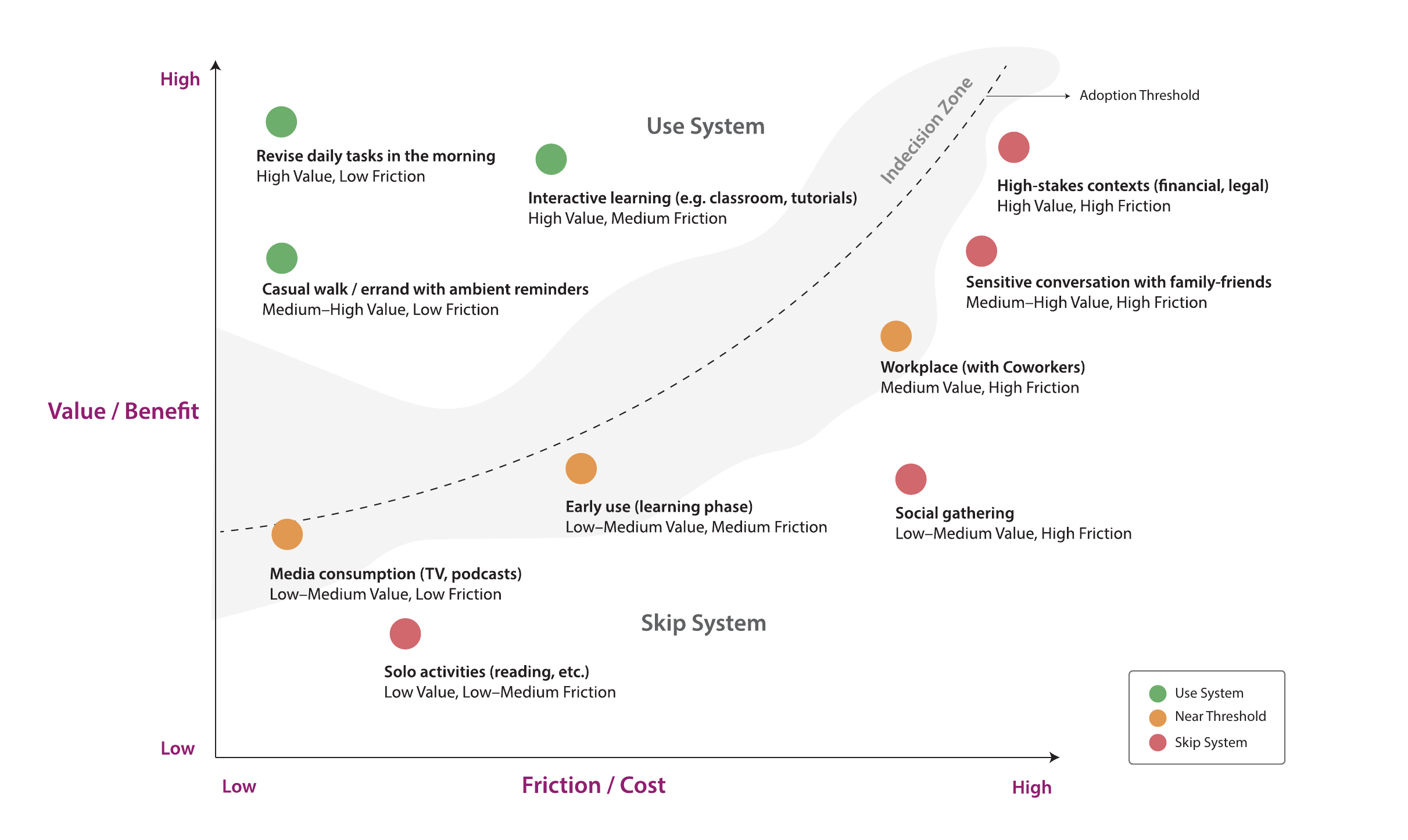}
    \caption{\textbf{Contextual Adoption Threshold in Everyday Use of \textsc{Memoro 2}.}
    Contexts participants described, placed by perceived value and friction. Positions and the dashed ``adoption threshold'' are approximate and reflect our interpretation of interview data, not measurements. Use was more likely above the line and avoided or deferred below it; in the shaded \textit{indecision zone}, use was situational.}
    
    \Description{A 2d graph showing the contextual adoption threshold, with y axis representing value and x axis representing the frictions. There are various circular points on the graph that represent different contexts where users navigate the tradeoff between the functional value and frictions associated with ubiquitous capture. There is a dashed line along the graph with an upward slope. The line illustrates a representative ``adoption threshold,'' reflecting our interpretation of interview data rather than a measured boundary. There is a greyed area around the boundary to show an indecision zone where participants could choose either way. Each point (context) is colored one of three colors signifying use, skip or near threshold.}
    \label{fig:adoption-threshold}
\end{figure*}

Using analysis of interaction logs, themes from participant interviews, and participant-led design suggestions for the system, we synthesized what emerged regarding always-on wearables using AI for memory support.

\subsection{Adopting Always-On Memory Support as a Context-Dependent Tradeoff}
\label{sec:adoption}

\begin{heroquote}
``At first it felt like it was watching us. Later it felt like it was helping us.'' (P1, 71 years old)
\end{heroquote}

Older-adult adoption of AI voice assistants is documented as selective, value-laden, identity-mediated, household-mediated, and recurrent~\cite{trajkova2020alexa, brewer2022alexa, pradhan2025not, soubutts2022amazon, bonilla2020older, sin2022does, upadhyay2023studying, mathur2025feels}. Older adults further resist technologies whose use re-categorizes them as in decline~\cite{caldeira2022stigma, kottl2021but, knowles2021harm}. Our 10-day deployment is consistent with these findings and adds two perspectives specific to always-on memory support, the value that drives adoption, and the handling of trust in a visible memory aid and the associated stigma. 

\subsubsection{Value Centered on Prospective Memory Support}

For several participants, the motivation to use \textsc{Memoro 2} was driven by concerns about prospective memory, remembering future intentions. Prospective memory failures are particularly consequential for older adults because they undermine functional independence~\cite{henry2004prospective, kliegel2016prospective}. \textsc{Memoro 2} proactively surfaced verbalized future intentions and let participants pull details back through conversational query, which could supplement offline prospective-memory training through implementation intentions in prior work with older adults~\cite{chan_prospecfit_2019}. Additionally, several participants used self-talk to take notes on future intentions, aligning with cognitive offloading as strategic delegation~\cite{risko2016cognitive}. Fig.~\ref{fig:adoption-threshold} summarizes this pattern as a contextual adoption threshold, a schematic interpretation of interview data rather than measured values: use clusters in low-friction routines (e.g., morning task revision, ambient reminders) and learning-loaded contexts. Skips concentrate in high-stakes settings, even when the value could be high, due to legal constraints. This extends to sensitive conversations as well, where, while memory lapses could be more critical, recording them comes with high social costs.

\subsubsection{Trust and Stigma in Memory-Aid Use}
\label{subsec:trust-stigma}

A memory aid captures and returns the user's own conversations, so trust forms differently than for a voice assistant. Voice-assistant trust has
been characterized through source credibility, transparency, and predictability \cite{bonilla2020older, sin2022does, mathur2025feels}, and
through the representational politics of voice and persona \cite{pradhan2025not}. For a memory aid, the source is the user's own remembered experience, and the system's account is read against that internal record. In our deployment, summaries that matched participants' recollections (P2) gave them a verifiable surface, and that surface extended to proactive reminders and recall answers, outputs the user cannot otherwise check. This way of building trust carries a risk: a system that matches someone's memory can confirm errors as well as correct them \cite{pat_2025}, and the contrast between in-situ helpfulness (94\%) and offline accuracy when answering (72\%), though measured on different question sets, is consistent with participants approving answers they had no way to check. Stigma operates alongside trust. Caldeira et al.\ \cite{caldeira2022stigma} capture older adults' stance toward assistive technology as ``I hope I never need one,'' which holds even when the technology is plainly useful because using it re-categorises the user, an effect HCI work on aging has cautioned against \cite{knowles2021harm, kottl2021but, vines2015ageold}. Two design choices in \textsc{Memoro 2} respond to this. The form factor is a commercially generic bone-conduction headset rather than a visibly assistive device (Fig. ~\ref{fig:teaser}); P2 carried the same logic forward in proposing smart glasses, since most older adults already wear glasses. Private audio prompts kept needing a reminder from becoming a public signal. For memory aids, the archive itself can be read as a ledger of need, so trust must be designed not only between user and system but in how use is read by others.

\subsection{Personal Memory Support Functioning as a Shared Social Tool}

\begin{heroquote}
``If I know I'm gonna run into them, I would say\ldots\ `just give me a summary of what we talked about last time'\ldots\ Next time I speak with them.'' (P5, 71 years old)
\end{heroquote}

Although \textsc{Memoro 2} was designed as a personal cognitive aid, its in-the-wild deployment enabled us to observe how it was embedded and reshaped participants' social fabric. Many field studies of AI assistants with older adults have examined stationary, home-based devices~\cite{chang2024longterm, jones2021lonelinessPVA, kim2021chb}. By following participants into daily life outside the home, this study reveals social dynamics that may go unseen in domestic deployments.


\subsubsection{Changes in Daily Speech during Interpersonal Interactions}

Prior wearable lifelogging work has examined whether capture changes the wearer's own behavior. Hoyle et al.\ report that most lifeloggers did not substantially modify their behavior under camera capture~\cite{hoyle_privacy_2014}.  In our audio-based deployment, the device's data was readable only by the user and stored locally, yet participants oriented to it as a third party. Early shifts toward formality (P2's ``newscaster,'' P6's ``King's English'') are better read as an accommodation to the user's anticipated future review than as performance for an outside observer, and these shifts declined over time. Where reflective informatics returns abstracted personal data such as steps or mood as the surface for self-examination \cite{baumer_reflections_2014, li_understanding_2011}, here the surface is conversation itself, and participants used it to recalibrate tone and humor. The capture also folds in interlocutors. Worry that partners would self-edit eased in close ties as the device drifted into shared use (P1) and lingered with peripheral contacts. This split has a structural cause. The wearer can listen back to the captured record and recalibrate accordingly. The partner has no such record, so any adjustments they make ease only as familiarity with the device or the wearer grows. Older adults prioritize their daily interaction with a few close people \cite{carstensen1999taking}, so the friction may be less apparent with longitudinal adoption of always-on systems.

\subsubsection{Aiding Social Connections Through Memory Support}
Our findings suggest a design space in which memory tools can support not only individual cognition but the ongoing work of maintaining relationships. Loneliness is a pervasive concern among older adults, with reduced social participation among its primary drivers~\cite{jones2021lonelinessPVA}. Although \textsc{Memoro 2} was built as a personal tool, it occasionally drifted into a joint household utility, a role also previously documented in shared, screen-based home devices~\cite{soubutts2022amazon}, but here arising from a wearable system. This shift shows that always-on systems can serve conversation partners who never use the system: the shared record becomes a referee for settling disagreements. The system also eased the emotional labor of social life. Forgetting something a close contact recently shared is among the most commonly reported everyday memory errors for older adults~\cite{ossher_everyday_2013}, and such lapses can carry real emotional weight in close relationships~\cite{stites_relationships_2018}. By reviewing details before sensitive or high-stakes conversations, such as visiting a grieving friend, participants reported feeling more confident and more present. By lowering these cognitive and emotional barriers, \textsc{Memoro 2} may help sustain the social connections that buffer against isolation.

\subsection{Designing Always-On for Adaptively-On Use}

\begin{heroquote}
``If I knew I could immediately delete it\ldots\ that would make a big difference.'' (P6, 65 years old)
\end{heroquote}

Section \ref{sec:adoption} showed adoption as moment-to-moment re-evaluation. The design question that follows is not whether to allow always-on capture, but how to support the active and adaptive regulation participants do throughout the day. Privacy-by-Design~\cite{cavoukian2009privacy} provided the baseline: participants did not find \textsc{Memoro 2} invasive, and understanding the data lifecycle at onboarding shaped initial willingness to engage. The mechanisms required for sustained use surfaced only once the system entered daily life (Section \ref{subsec:rq3}), and they group into two design directions: bystander coordination as continuous practice, and selective capture supported by adjusted visibility and post-capture controls.

\subsubsection{Bystander Coordination as Continuous Practice}
Bystander consent required ongoing negotiation, particularly in group settings where new people entered mid-capture. Consent norms are relationship- and context-specific, so they surface at the moment of capture, not before it. Handling varied by relationship: easier with established contacts, where trust reduced the social cost of disclosure; harder with new encounters, mirroring Price et al.'s finding that lifeloggers apply different norms to insiders versus outsiders ~\cite{price2017logging}. Management improved with the use, where participants pre-notified contacts electronically, and frequent interlocutors began recognizing the device on their own. Prior work treats bystander consent mainly as a pre-deployment design challenge~\cite{denning2014bystanders, hoyle_privacy_2014, ohagan2023pet}; recent framings call for context-adaptive pathways between wearer and bystander~\cite{wang2026mindgap}. Our deployment documents how that competency develops in situ, refined by prior context and familiarity with the interlocutor. Under contextual integrity~\cite{nissenbaum2004privacy}, these strategies restore norms that continuous capture disrupts. Systems should treat consent as an ongoing practice, tracking who and in what context people have been informed, surfacing the dynamic capture state and reasoning to bystanders, and supporting pre-notification, rather than treating it as a one-time gate.

\subsubsection{Selective Capture and Adjustable Visibility}
Participants sought control over both how visible the system was and what it retained, pointing to a continuum rather than a binary on/off, reinforcing Malkin et al.'s~\cite{malkin_privacy_2020} call for granular controls in mobile wearable contexts. One participant's ``temperature dial'' suggestion captures the requirement: quiet in social spaces, explicit when reliability mattered. Beyond visibility, concerns extended to what persists in the record, adding a post-capture dimension to the continuous re-evaluation in Section \ref{sec:adoption}. Participants worried about recordings traveling beyond their original setting, context collapse, where content captured under one set of norms is later read under another. Their emphasis on immediate deletion (P6) indicates near-real-time control over retention, consistent with older-adult-specific calls for stronger retention control on always-listening devices~\cite{lau_alexa_2018, pradhan2020iva, spangler2021privacy}. Prior work frames privacy for always-listening devices around surveillance and data exposure~\cite{dunbar2021listening}. For older adults, a further concern is that a memory aid's archive could be read as evidence of decline rather than capability extending documented concerns about stigma and identity~\cite{caldeira2022stigma, kottl2021but}. Suggestions converge on rolling buffers, discard-by-default policies, and faceted filters. Two design distinctions follow: adjusted visibility regulates moment-to-moment visibility of system state; post-capture controls regulate what the record contains about the user, distinct from data protection, which limits who else can see the record. The distinction matters where the record itself can be read as a ledger about its subject.

\section{Limitations and Future Work}

Our deployment reveals both the promises and constraints of bringing AI memory agents into the everyday lives of older adults. We highlight key limitations of our study and design, and outline directions for future work.

First, participants noted challenges with carrying a separate phone and Bluetooth headset, which required extra effort to connect and charge; P3 found coordinating multiple devices difficult. At the system level, speaker identification was a limitation that restricted speaker-based queries. Reminders were also limited to time-based scheduling and did not support event-based triggers, such as reminding to take keys before leaving. Regarding technical evaluation, privacy-preserving local storage precluded ground-truth benchmarking on participant data, so we paired AMV's in-situ helpfulness ratings with the offline component validation of Section~\ref{subsec:offline_validation}. That evaluation surfaced concrete directions for future systems, most prominently cross-session deduplication of extracted commitments, and its scope is deliberately matched to the probe's ten-day deployment.

Second, this study likely surfaced interactions that were useful within a ten-day span. Longer deployments are necessary to examine how captured memories provide value over months or years. Participants suggested that longer use would likely reveal new ways to interact with the system and might increase the use of categories like nostalgia and relationships. Our recruitment was restricted to urban settings, which may not reflect routines in rural areas with different community norms, and professional environments such as hospitals can constrain use. Future deployments should span more diverse demographic, social, and professional contexts.

Third, while our participants did not report cognitive impairments, caregiving participants noted the potential value for those with dementia. Future studies should adapt interaction models for cognitive accessibility, such as simplified interfaces or caregiver-mediated control \cite{xu2026remiassist}. Social activity levels also shape adoption. For participants who live alone, fewer social encounters may reduce the perceived value of the system. The sample skewed male (8 of 10), and 6 participants had prior exposure to voice assistants. Findings reflect tech-fluent older adults, who would likely represent early adopters; less-exposed or older populations may engage differently, and in-situ helpfulness ratings are voluntary self-reports that may skew positive. Legal and ethical restrictions in sensitive contexts further underscore the need for strategies such as opt-in consent or selective suppression of capture. Finally, growing work shows the possibility of false memories induced by AI systems; these need to be carefully considered in longitudinal studies \cite{pat_2025, chan2024conversationalaipoweredlarge}. Accounts of bystander and partner reactions come from participants' own reports; bystanders were not directly interviewed.

\section{Conclusion}

Our 10-day in-the-wild deployment shows that always-on AI memory support was used most when prospective memory needs were concrete and when its outputs matched participants' own recollection. Participants weighed benefit against social cost continuously, mid-conversation and after, and the system folded into existing morning planning through reminders and end-of-day review through summaries. Capture initially formalized participants' own speech and raised concerns that partners would self-edit, both of which eased over time, though the worry lingered with peripheral contacts. The personal aid sometimes became a shared household tool for settling disagreements, preparing for sensitive conversations, and following up on social obligations. Two design directions follow: selective capture, supported by adjustable visibility for moment-to-moment regulation and post-capture controls over what the record retains; and consent management for bystanders as an ongoing practice. These findings ground always-on AI memory support in the everyday realities of older adults' lives.

\begin{acks}
This work was supported by an MIT Aging Brain Initiative Seed Fund Award and Google.org.
\end{acks}

\bibliographystyle{ACM-Reference-Format}
\bibliography{references, proactive_ai, memory_assistants, references_update}

%
\appendix
\input{meta_appendix}

\end{document}

%% file: meta_appendix.tex

\section{Methodological Transparency \& Reproducibility Appendix (META)}
\label{appendix:meta}

This appendix consolidates the implementation, deployment, and analysis details required to repeat or replicate the study. Verbatim LLM prompt text, the design-rationale quote table, and the qualitative codebook are provided as supplementary materials; the items below are pointers, parameters, and instruments.


\subsection{Hardware and Software Stack}
\label{appendix:meta:stack}

Each participant received a Samsung Galaxy A35 running the \textsc{MemHelp} Android application, a Shokz OpenRun Pro~2 bone-conduction headset, and an optional ``I'm recording'' wearable badge. The mobile application performed on-device Voice Activity Detection (VAD), local AES-256-GCM encryption with device-keystore keys, and reminder scheduling via Android \texttt{WorkManager}. Server-side processing used the OpenAI \texttt{gpt-4o-realtime-preview-2024-12-17} model for the recall and reminder agents, GPT-4o for chunking, summarization, significance scoring, and interaction classification, \texttt{text-embedding-3-small} for chunk vectorization, and the ElevenLabs API~\cite{ElevenLabs} for transcription, diarization, and text-to-speech.

\subsection{Pipeline Parameters}
\label{appendix:meta:params}

\textit{Capture.} The on-device VAD operates as a four-state machine: a 5-second rolling pre-roll buffer, onset detection on first speech, segmentation into clips of up to 2~minutes, and a 30-second silence timeout that closes the conversation and returns the system to idle.

\textit{Memory encoding.} Each transcript is segmented into 3--8 chunks per hour at topic boundaries; each chunk receives a 150--200 word summary and an integer significance score in $[0,10]$ from GPT-4o. Summaries are vectorized with \texttt{text-embedding-3-small} (384-dimensional embeddings) for semantic search.

\textit{Recall ranking.} Candidate chunks are ranked by $P = \alpha \cdot \text{Recency} + \beta \cdot \text{Significance} + \gamma \cdot \text{Relevance}$ with $\alpha = -0.01$ (recency decay), $\beta = 0.02$ (importance), $\gamma = 1.0$ (semantic similarity). The recall agent is limited to five tool calls per query, after which a fixed fallback response is returned.

\textit{Reminders.} Informal time references in conversation are converted to absolute timestamps (MM-DD-YYYY,~HH:MM:SS). Scheduled times are constrained to a 6~AM--11~PM window and to within 14 days of the study end; deadline-style reminders fire 30 minutes prior to the deadline.

\subsection{Storage and Privacy Implementation}
\label{appendix:meta:privacy}

The deployment instantiated Cavoukian's Privacy-by-Design principles~\cite{cavoukian2009privacy} via four mechanisms: (i)~AES-256-GCM encryption at rest with device-keystore keys; (ii)~zero-data-retention contracts with the LLM provider for the duration of the study, with no audio or transcript content retained, logged for abuse monitoring, or used for model training; (iii)~deletion of raw audio on receipt of the transcript, with raw audio never persisted server-side; (iv)~a mobile dashboard exposing per-conversation, per-recall, and per-reminder deletion plus a single-action ``erase all'' control.

\subsection{Recruitment and Deployment Protocol}
\label{appendix:meta:protocol}

\textit{Recruitment.} Participants were recruited via email announcements to community members and in-person outreach at local senior centers in an urban area. Eligibility required age $\geq 60$, no hearing or vision impairment that would interfere with a wearable/voice interface, no formally diagnosed cognitive impairment, and self-described openness to everyday technology use. The screening survey verifying these criteria appears as Exhibit~\ref{appendix:meta:surveys} below.

\textit{Onboarding (60 minutes, on-campus).} (1) Orientation \& purpose (5 min); (2) feature walkthrough (5 min); (3) bystander notification practice (5 min); (4) hands-on setup including voice signature recording (5 min); (5) practice time with a semi-structured conversation using the trigger phrase ``hey navi'' (30 min). The exit interview was 30 minutes; check-in phone interviews of 15--20 minutes were conducted every other day during the deployment.

\textit{Compensation.} \$150 on study completion; \$50 additional for participants who provided in-app feedback ratings on memory interactions. The compensation level reflected lab visits, daily device use, and 15--20 minute interviews every other day.

\textit{Evaluation protocol.} We followed the Automated Memory Validation (AMV) protocol~\cite{elagroudy_pixel_2025}: participants rated each recall and reminder via in-app thumbs up/down and optional comments. Researchers received only the validation labels, never the underlying audio or transcripts. AMV produces a privacy-preserving signal of perceived helpfulness on returned outputs and cannot count silent failures.

\textit{IRB.} The protocol was approved by the institutional review board prior to recruitment.

\subsection{Coding Procedure}
\label{appendix:meta:coding}

Interviews were transcribed and analyzed via reflexive thematic analysis with a hybrid inductive--deductive approach across four stages: (1) familiarization with transcripts and recordings, linking excerpts to research questions; (2) integration with participant-level usage summaries from system logs; (3) reflexive coding with iterative refinement through independent coding and negotiated agreement; (4) theme development and validation against the full dataset. The full codebook is provided as supplementary material (\texttt{CodeBook-Supplementary Material.xlsx}).

\subsection{Supplementary Materials and Exhibits}
\label{appendix:meta:pointers}

\textit{Supplementary materials} (separate file, not part of the manuscript): verbatim LLM prompts for chunking, daily summarization, recall query construction, and recall/reminder classification; the design-rationale table mapping participant quotes to design choices; and the qualitative codebook. \textit{In-META exhibits below}: study survey instruments (Exhibit~\ref{appendix:meta:surveys}), interview scripts (Exhibit~\ref{appendix:meta:interviews}), and the per-category interaction count table (Exhibit~\ref{appendix:meta:categorization}). The \textsc{MemHelp} mobile application code will be released as open source upon publication (repository link withheld for review anonymity);


\subsection{Survey Instruments}
\label{appendix:meta:surveys}

\subsubsection*{Recruitment Survey}
\begin{enumerate}\setlength{\itemsep}{0pt}\small
\item Name.
\item Age.
\item Preferred email.
\item Gender.
\item Occupation.
\item Diagnosed neurological or cognitive condition (e.g., Alzheimer's, schizophrenia, stroke-related impairment)?
\item Daily-life memory difficulties (e.g., recent conversations, doctors' appointments)?
\item If yes, frequency.
\item Hearing difficulties or hearing-aid use?
\item Hearing aid brand and model (if applicable).
\item Visual impairment?
\item Native English speaker?
\item Speech impairment?
\item Smartphone use frequency.
\item Confidence with new smartphone apps.
\item Prior AI-assistant use (Alexa, Siri, etc.).
\item Existing memory-aid technology use (which ones?).
\item Comfort wearing an open-ear headset throughout the day.
\item Comfort informing others that conversations are being recorded.
\item Significant daily periods when audio recording is not possible.
\item Self-description as open to exploring new technology.
\end{enumerate}

\subsubsection*{Needs-Finding Survey}
\begin{enumerate}\setlength{\itemsep}{0pt}\small
\item Year of birth.
\item Familiarity with AI-powered devices/assistants.
\item Likelihood of use --- summarizing a meeting or interaction.
\item Likelihood of use --- finding the name of someone you met.
\item Likelihood of use --- reminder about a discussed task (e.g., a promise).
\item Likelihood of use --- reminder about something told to the system (e.g., an appointment).
\item Likelihood of use --- recall in the middle of a conversation.
\item Scenario 1, Yesterday recap: phrasing to ask the system about yesterday.
\item Scenario 2, Son's update: phrasing to recall what the user's son said about work.
\item Scenario 3, Medication timing: phrasing to query medication time.
\item Scenario 4, Lunch reminder: phrasing for tomorrow's lunch reminder.
\item Scenario 5, Friend review: phrasing to recall a friend's recent updates before a visit.
\item In-conversation name recall: phrasing during a chat about new neighbors.
\item Open-ended: other use cases the participant envisions.
\end{enumerate}

\subsection{Interview Scripts}
\label{appendix:meta:interviews}

\subsubsection*{Phone Check-In Script (every other day, 15--20 min)}
Opening: ``This call is part of a study about how people interact with memory and reminders in daily life, and how a new tool called \textsc{MemHelp} might support that. This phone call will be recorded, is that okay?''
\begin{enumerate}\setlength{\itemsep}{0pt}\small
\item Experience using \textsc{MemHelp} so far.
\item Times the participant did not want to use it.
\item Feeling of having a device listen during the day.
\item Kinds of moments or things asked to be remembered.
\item Recall attempts and how they felt.
\item Helpful/unhelpful examples for reminders, recalls, conversations.
\item Examples of when it did and did not work.
\item Behavioral changes made to use \textsc{MemHelp}.
\end{enumerate}

\subsubsection*{Exit Interview Script (30 min)}
Final-interview prompts iterated based on usage patterns from prior check-ins.
\begin{enumerate}\setlength{\itemsep}{0pt}\small
\item \textbf{Introduction:} imagine \textsc{MemHelp} fully developed and part of everyday life.
\item \textbf{Daily summaries:} what makes the day-recap feel powerful or helpful.
\item \textbf{Confidence in remembering:} effect on self-confidence about memory.
\item \textbf{Information presentation:} how recalls and reminders should be ordered/displayed.
\item \textbf{Form factor:} effect of removing the headset and running in the background.
\item \textbf{Integration:} value vs. complexity of syncing with Outlook, Google Calendar, etc.
\item \textbf{Boundaries:} conversation types where the system should not listen.
\item \textbf{Relationships:} effect of having a ``ground truth'' record on relationships.
\item \textbf{Future use:} role of the system 5 years out.
\item \textbf{Potential users:} who would benefit most from daily use.
\item \textbf{Explaining benefits:} how the participant would describe \textsc{MemHelp} to others.
\item \textbf{Additional features:} other capabilities the participant would want.
\end{enumerate}

\subsection{Interaction Categorization}
\label{appendix:meta:categorization}

\begin{table}[h]
\centering\small
\caption{Recall and reminder counts by interaction category across the 10-day deployment.}
\label{tab:meta-category-counts}
\begin{tabular}{|l|c|c|}
\hline
\textbf{Category} & \textbf{Reminders} & \textbf{Recalls} \\
\hline
Tasks \& Chores & 109 & 61 \\
Appointments \& Schedules & 55 & 40 \\
Health \& Wellness & 15 & 11 \\
Social Plans \& Shared Events & 11 & 5 \\
Names \& Relationships & 10 & 41 \\
Instructions \& How-Tos & 9 & 3 \\
Places \& Travel & 9 & 18 \\
Promises \& Commitments & 7 & 2 \\
News \& Current Events & 3 & 1 \\
Factual Information & 1 & 18 \\
None of the above & 0 & 28 \\
Personal Memories \& Nostalgia & 0 & 7 \\
\hline
\end{tabular}
\end{table}

\subsection{Offline Validation: Corpus Construction and Additional Conditions}
\label{appendix:meta:offline}

\textit{Corpus.} Conversation seeds (personas, story arcs, commitment categories, distractor types) were authored to mirror the style of the deployment conversations; gold labels (commitment spans, absolute event times, acceptable scheduling windows, QA evidence turns) were annotated at authoring time and machine-validated for consistency. Question timing follows the in-situ protocol of Section~\ref{subsec:offline_validation}: each question runs against the memory store as of its asking time, with the reminder database replayed to that moment.

\textit{Scoring.} LLM judge (GPT-4o, temperature 0) with lenient paraphrase matching calibrated for spoken answers; reminder-commitment matching via embedding candidates with judge confirmation; all raw generations, tool calls, and judgments are retained.

\textit{Human verification.} Three crowd raters (Prolific), blinded to the judge's labels and to each other, re-judged all 50 recall answers on a five-point scale (correct / correct but incomplete / incorrect / no answer / can't tell) and a stratified 19\% sample of reminder-commitment pairs; every rater passed all six embedded attention checks. When the five-point labels are collapsed to the judge's binary decision, the rater majority matched the judge on 33 of 35 answerable questions (94\%) and all 15 unanswerable ones, for 48 of 50 (96\%) overall, above the roughly 80\% human-judge agreement reported for LLM judges on open-ended tasks~\cite{zheng2023judging}. Raters were unanimous on 25 of 35 items (Fleiss $\kappa$ = 0.59 on the five-point labels, 0.72 after the binary collapse). In both disagreements the raters were more lenient than the judge: they unanimously accepted two answers the judge had failed for omitted details. No rater majority rated as factually wrong any answer the judge had passed, so Table~\ref{tab:offline_validation} keeps the judge's stricter labels. Raters confirmed 9 of 10 sampled reminder matches (one majority ``can't tell'') and unanimously judged that all five duplicate-flagged reminders repeated commitments that were already captured, consistent with the duplication finding in Section~\ref{subsec:offline_validation}. A pilot rater, excluded from analysis, exposed an ambiguity in the incompleteness option, which was anchored to the reference rather than the question; the rubric was corrected before the main run.

\textit{Corpus face validity.} Raters also scored three ten-turn excerpts for naturalness, plausibility, and coherence on five-point scales; median ratings were 4 or higher on every scale for every excerpt.
\textit{Additional condition (transparency).} An earlier, deliberately adversarial version of the corpus (near-miss unanswerable questions, injected ASR noise, and questions asked only after the corpus week) yields lower recall accuracy (51\% answerable, 87\% refusal) and is reported in the supplementary material as a stress condition.